\def\year{2018}\relax
\documentclass[letterpaper]{article} 
\usepackage{aaai18}  
\usepackage{times}  
\usepackage{helvet}  
\usepackage{courier}  
\usepackage{url}  
\usepackage{graphicx}  
\usepackage{algorithmicx}
\usepackage{algorithm}
\usepackage{algpseudocode}
\usepackage{amsmath}
\usepackage{amssymb}
\usepackage{booktabs} 
\usepackage{color}
\usepackage{enumitem} 
\usepackage{graphicx}
\usepackage{multirow}
\usepackage{subcaption}
\usepackage{url}

\renewcommand*\Call[2]{\textproc{#1}(#2)}

\begin{document}
\title{HARP: Hierarchical Representation Learning for Networks}
\author{Haochen Chen\\Stony Brook University\\haocchen@cs.stonybrook.edu
\And Bryan Perozzi\\Google Research\\bperozzi@acm.org
\AND Yifan Hu\\Yahoo! Research\\yifanhu@oath.com
\And Steven Skiena\\Stony Brook University\\skiena@cs.stonybrook.edu}
\maketitle

\newcommand{\ouralgorithm}{HARP}
\newcommand{\ourdw}{HARP(DW)}
\newcommand{\ourline}{HARP(LINE)}
\newcommand{\ourntv}{HARP(N2V)}

%
%
%


\begin{abstract}

We present \ouralgorithm, a novel method for learning low dimensional embeddings of a graph's nodes which preserves higher-order structural features. 
Our proposed method achieves this by compressing the input graph prior to embedding it, effectively avoiding troublesome embedding configurations (i.e. local minima) which can pose problems to non-convex optimization.
 
\ouralgorithm\ works by finding a smaller graph which approximates the global structure of its input.
This simplified graph is used to learn a set of initial representations, which serve as good initializations for learning representations in the original, detailed graph.
We inductively extend this idea, by decomposing a graph in a series of levels, and then embed the hierarchy of graphs from the coarsest one to the original graph. 

\ouralgorithm\ is a general meta-strategy to improve \emph{all} of the state-of-the-art neural algorithms for embedding graphs, including \emph{DeepWalk}, \emph{LINE}, and \emph{Node2vec}. Indeed, we demonstrate that applying \ouralgorithm's hierarchical paradigm yields improved implementations for all three of these methods, as evaluated on classification tasks on real-world graphs such as \emph{DBLP}, \emph{BlogCatalog}, and \emph{CiteSeer}, where we achieve a performance gain over the original implementations by up to 14\% Macro F1.

\end{abstract}

\section{Introduction} 

\label{Introduction}
\begin{figure}[!t]
	\centering
	\begin{subfigure}[b]{.32\linewidth}
		\includegraphics[width=\linewidth]{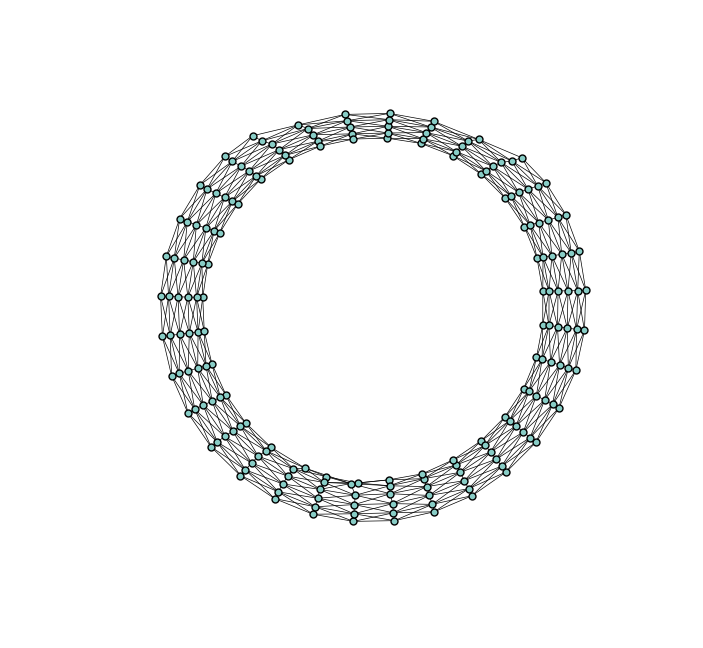}
		\caption{Can\_187}
		\label{fig:can_187_sfdp}
	\end{subfigure}
	\begin{subfigure}[b]{.32\linewidth}
		\includegraphics[width=\linewidth]{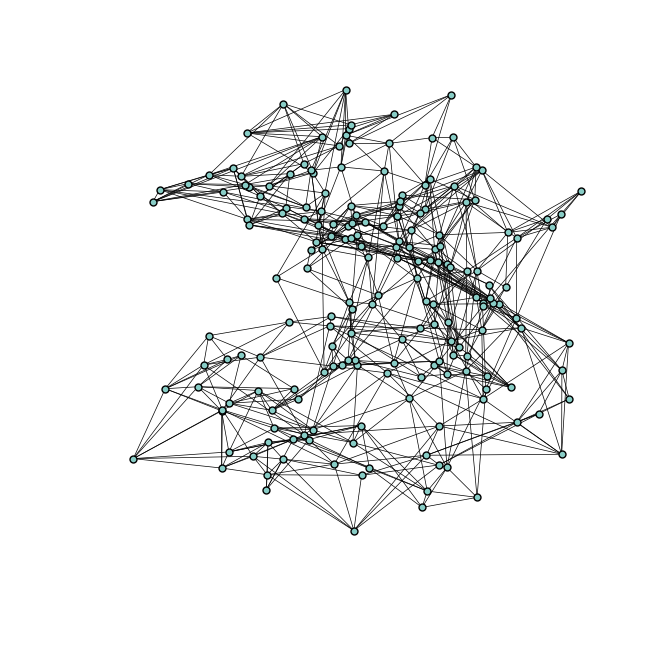}
		\caption{LINE}
		\label{fig:can_187_line}
	\end{subfigure}
	\begin{subfigure}[b]{.32\linewidth}
		\includegraphics[width=\linewidth]{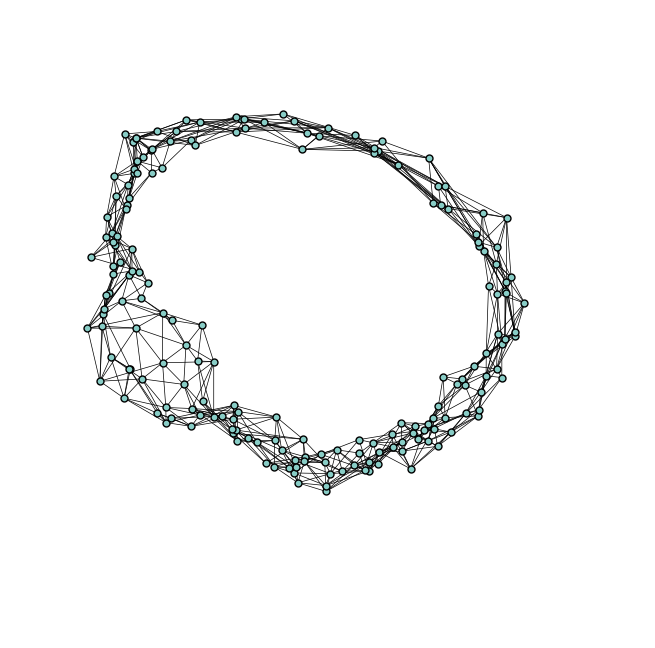}
		\caption{\ouralgorithm}
		\label{fig:can_187_line_gc}
	\end{subfigure}
	
	\begin{subfigure}[b]{.32\linewidth}
		\includegraphics[width=\linewidth]{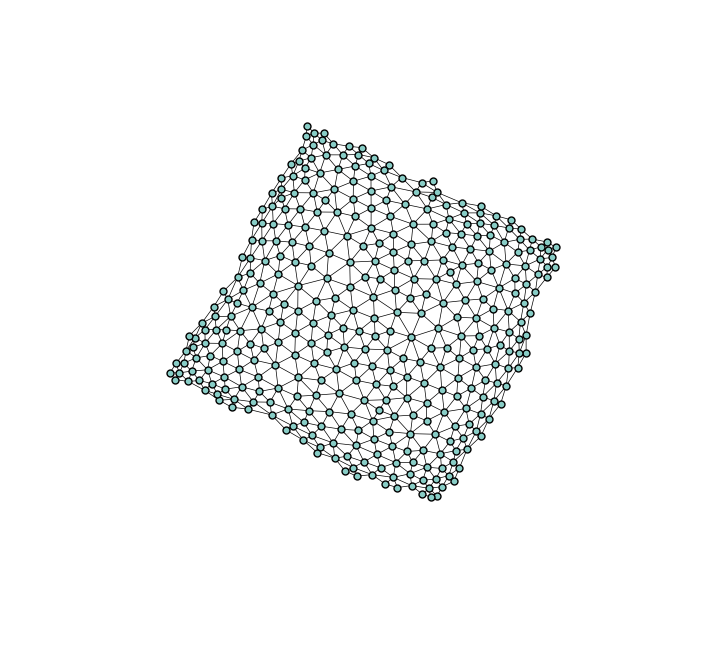}
		\caption{Poisson 2D}\label{fig:poisson_2d_sfdp}
	\end{subfigure}
	\begin{subfigure}[b]{.32\linewidth}
		\includegraphics[width=\linewidth]{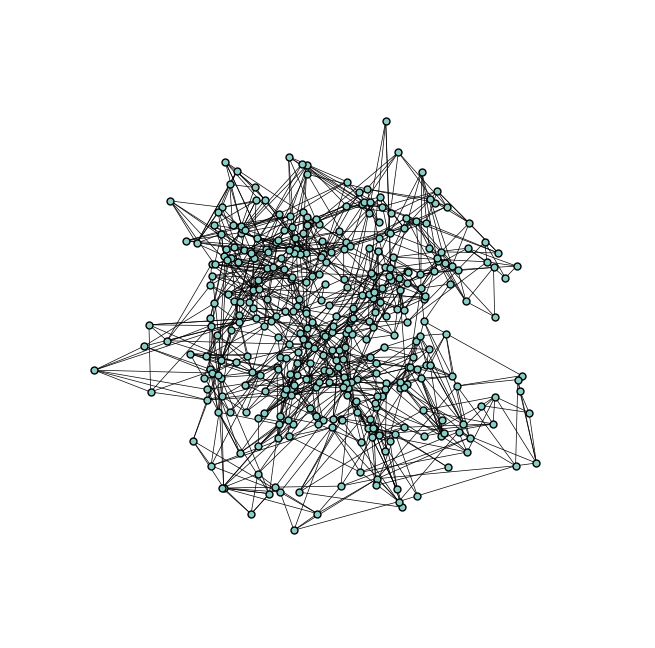}
		\caption{LINE}\label{fig:poisson_2d_line}
	\end{subfigure}
	\begin{subfigure}[b]{.32\linewidth}
		\includegraphics[width=\linewidth]{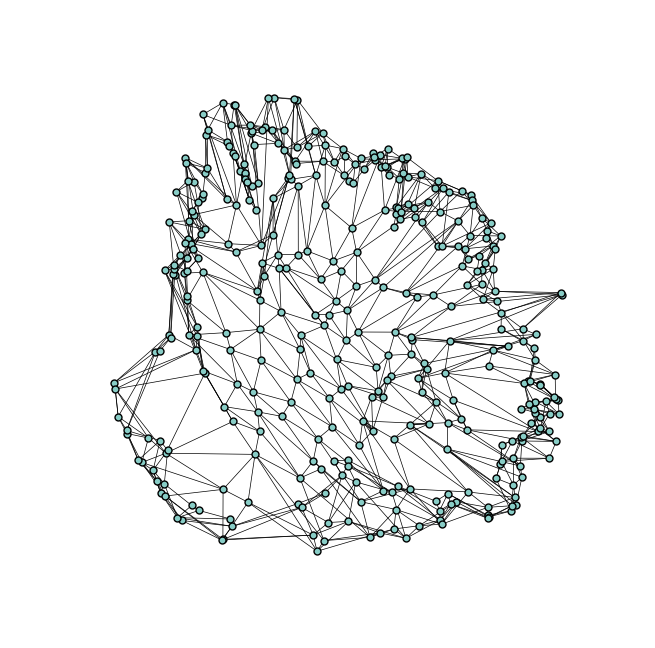}
		\caption{\ouralgorithm}\label{fig:poisson_2d_line_gc}
	\end{subfigure}
	\caption{Comparison of two-dimensional embeddings from LINE and our proposed method,
	for two distinct graphs.
	Observe how \ouralgorithm's embedding better preserves the higher order structure of a ring and a plane.}
	~\label{fig:graph_drawing_comp}
\end{figure}

\noindent
From social networks to the World Wide Web, graphs are a ubiquitous way to organize a diverse set of real-world information.
Given a network's structure, it is often desirable to predict missing information (frequently called \textit{attributes} or \textit{labels}) associated with each node in the graph.
This missing information can represent a variety of aspects of the data -- for example, on a social network they could represent the communities a person belongs to, or the categories of a document's content on the web.

Because many information networks can contain billions of nodes and edges, it can be intractable to perform complex inference procedures on the entire network. 
One technique which has been proposed to address this problem is \textit{dimensionality reduction}.
The central idea is to find a mapping function which converts each node in the graph to a low-dimensional latent representation.
These representations can then be used as features for common tasks on graphs such as multi-label classification, clustering, and link prediction.

Traditional methods for graph dimensionality reduction \cite{belkin2001laplacian,roweis2000nonlinear,tenenbaum2000global} perform well on small graphs.
However, the time complexity of these methods are at least quadratic in the number of graph nodes, makes them impossible to run on large-scale networks. 

A recent advancement in graph representation learning, DeepWalk \cite{perozzi2014deepwalk} proposed online learning methods using neural networks to address this scalability limitation.
Much work has since followed \cite{cao2015grarep,node2vec-kdd2016,walklets,tang2015line}.  
These neural network-based methods have proven both highly scalable and performant, achieving strong results on classification and link prediction tasks in large networks.

Despite their success, all these methods have several shared weaknesses.
Firstly, they are all local approaches -- limited to the structure immediately around a node.
DeepWalk \cite{perozzi2014deepwalk} and Node2vec \cite{node2vec-kdd2016} adopt short random walks to explore the local neighborhoods of nodes,
while LINE \cite{tang2015line} is concerned with even closer relationships (nodes at most two hops away).
This focus on local structure implicitly ignores long-distance global relationships, and the learned representations can fail to uncover important global structural patterns.
Secondly, they all rely on a non-convex optimization goal solved using stochastic gradient descent \cite{goldberg2014word2vec,mikolov2013distributed} which can become stuck in a local minima (e.g.\ perhaps as a result of a poor initialization).
In other words, all previously proposed techniques for graph representation learning can accidentally learn embedding configurations which disregard important structural features of their input graph.

In this work, we propose \emph{\ouralgorithm}, a meta strategy for embedding graph datasets which preserves higher-order structural features.
\emph{\ouralgorithm} recursively coalesces the nodes and edges in the original graph to get a series of successively smaller graphs with similar structure.
These coalesced graphs, each with a different granularity, provide us a view of the original graph's global structure.
Starting from the most simplified form, 
each graph is used to learn a set of initial representations which serve as good initializations for embedding the next, more detailed graph.
This process is repeated until we get an embedding for each node in the original graph.

We illustrate the effectiveness of this multilevel paradigm in Figure \ref{fig:graph_drawing_comp}, by visualizing the two-dimension embeddings from an existing method (\emph{LINE}~\cite{tang2015line}) and our improvement to it, \emph{\ourline}.
Each of the small graphs we consider has an obvious global structure (that of a ring (\ref{fig:can_187_sfdp}) and a grid (\ref{fig:poisson_2d_sfdp})) which is easily exposed by a force direced layout~\cite{hu2005efficient}.
The center figures represent the two-dimensional embedding obtained by LINE for the ring (\ref{fig:can_187_line}) and grid (\ref{fig:poisson_2d_line}).  In these embeddings, the global structure is lost (i.e.\  that is, the ring and plane are unidentifiable).
However, the embeddings produced by using our meta-strategy to improve LINE (right) clearly capture both the local and global structure of the given graphs (\ref{fig:can_187_line_gc}, \ref{fig:poisson_2d_line_gc}).

Our contributions are the following:
\begin{itemize}
\item \textbf{New Representation Learning Paradigm.}
We propose \emph{\ouralgorithm}, a novel multilevel paradigm for graph representation which seamlessly blends ideas from the graph drawing \cite{fruchterman1991graph} and graph representation learning \cite{perozzi2014deepwalk,tang2015line,node2vec-kdd2016} communities to build substantially better graph embeddings.

\item \textbf{Improved Optimization Primitives.}
We demonstrate that our approach leads to improved implementations of \textbf{all} state-of-the-art graph representation learning methods, namely \emph{DeepWalk} (DW), \emph{LINE} and \emph{Node2vec} (N2V).
Our improvements on these popular methods for learning latent representations illustrate the broad applicability of our hierarchical approach. 

\item \textbf{Better Embeddings for Downstream Tasks.}
We demonstrate that \emph{\ourdw}, \emph{\ourline} and \emph{\ourntv} embeddings
consistently outperform the originals on classification tasks on several real-world networks, with improvements as large as 14\% Macro $F_1$. 
\end{itemize}

\section{Problem Formulation}
We desire to learn latent representations of nodes in a graph.
Formally, let $G = (V, E)$ be a graph, where $V$ is the set of nodes and $E$ is the set of edges.
The goal of graph representation learning is to develop a mapping function $\Phi: V \mapsto \mathbb{R}^{|V| \times d}, d \ll |V|$.
This mapping $\Phi$ defines the latent representation (or \emph{embedding}) of each node $v \in V$.
Popular methods for learning the parameters of $\Phi$~\cite{perozzi2014deepwalk,tang2015line,node2vec-kdd2016} suffer from two main disadvantages:
(1) higher-order graph structural information is not modeled, and 
(2) their stochastic optimization can fall victim to poor initialization.

In light of these difficulties, we introduce the \emph{hierarchical representation learning} problem for graphs.
At its core, we seek to find a graph,  $G_s = (V_s, E_s)$ which captures the essential structure of $G$, but is smaller than our original (i.e.\ $|V_s| << |V|$, $|E_s| << |E|$).
It is likely that $G_s$ will be easier to embed for two reasons.
First, there are many less pairwise relationships ($|V_s|^2$ versus $|V|^2$) which can be expressed in the space.
As the sample space shrinks, there is less variation in training examples -- this can yield a smoother objective function which is easier to optimize.
Second, the diameter of $G_s$ may be smaller than $G$, so algorithms with a local focus can exploit the graph's global structure.

In summary, we define the hierarchical representation learning problem in graphs as follows:
\noindent
\begin{itemize}
	\setlength{\itemsep}{-1.0\itemsep}
	\item[]\hspace{-0.25in}
	{\bf Given} a large graph $G(V,E)$ and a function $f$, which embeds $G$ using initialization $\theta$, $f: G \times \theta \mapsto \Phi_G$,
	
	\item[]\hspace{-0.25in}
	{\bf Simplify} $G$ to a series of successively smaller graphs $G_{0} \hdots G_{L}$,
	
	\item[]\hspace{-0.25in}
	{\bf Learn} a coarse embedding $\Phi_{G_L} = f(G_L, \emptyset)$, 
	
	\item[]\hspace{-0.25in}
	{\bf Refine} the coarse embedding into $\Phi_{G}$ by iteratively applying $\Phi_{G_i} = f(G_i, \Phi_{G_{i+1}}), 0 \leq i < L$.
\end{itemize}

\section{Method}
\label{HARP}

Here we present our hierarchical paradigm for graph representation learning.
After discussing the method in general, we present a structure-preserving algorithm for its most crucial step, graph coarsening.


\begin{figure*}[t]
\centering
\begin{subfigure}[b]{.32\linewidth}
	\centering
\includegraphics[width=0.9\linewidth]{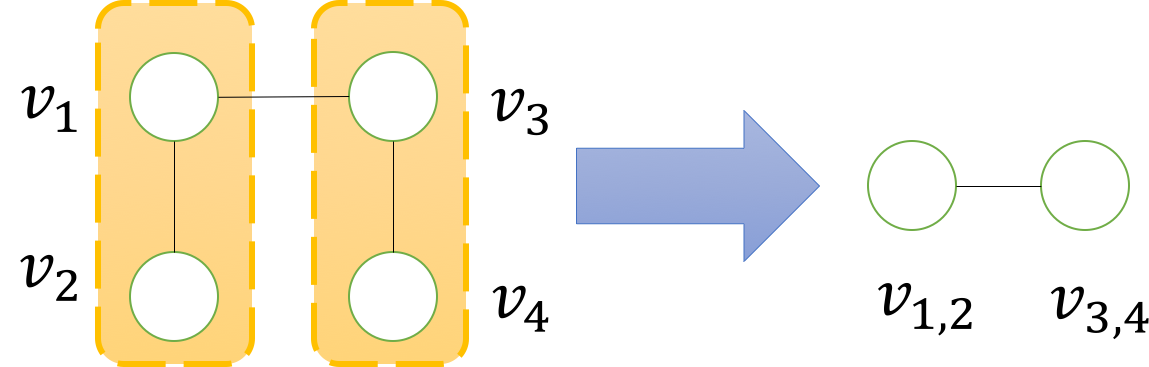}
\caption{Edge Collapsing.}
\label{fig:edge-collapsing}
\end{subfigure}
\begin{subfigure}[b]{.32\linewidth}
	\centering
\includegraphics[width=0.85\linewidth]{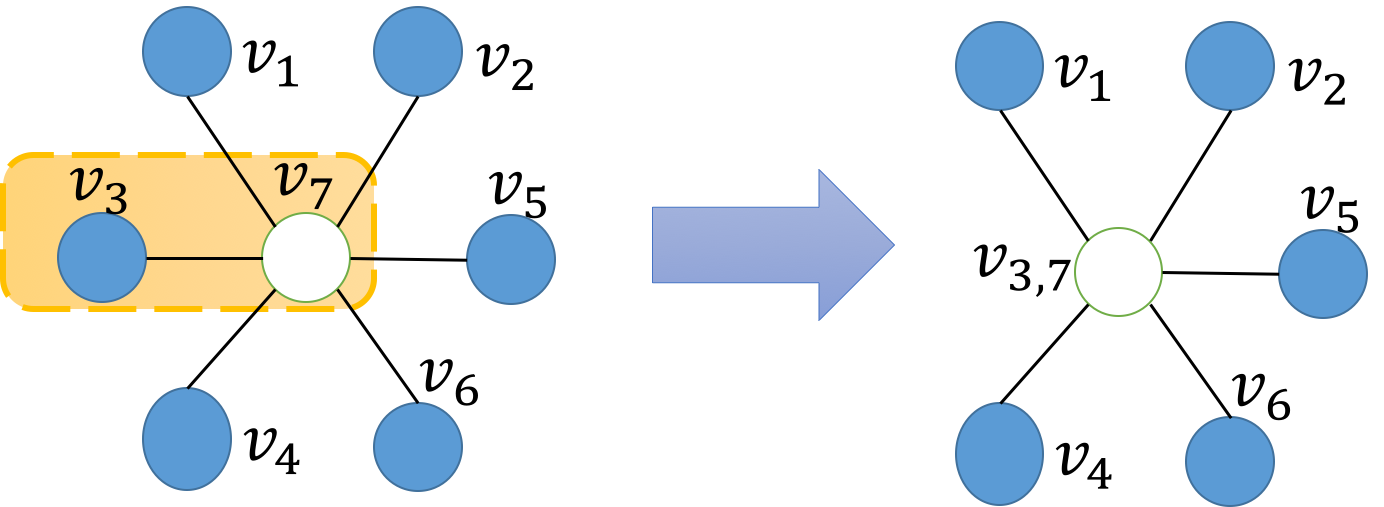}
\caption{Edge Collapsing fails to collapse stars.}
\label{fig:edge-collapsing-fail}
\end{subfigure}
\begin{subfigure}[b]{.32\linewidth}
	\centering
\includegraphics[width=0.85\linewidth]{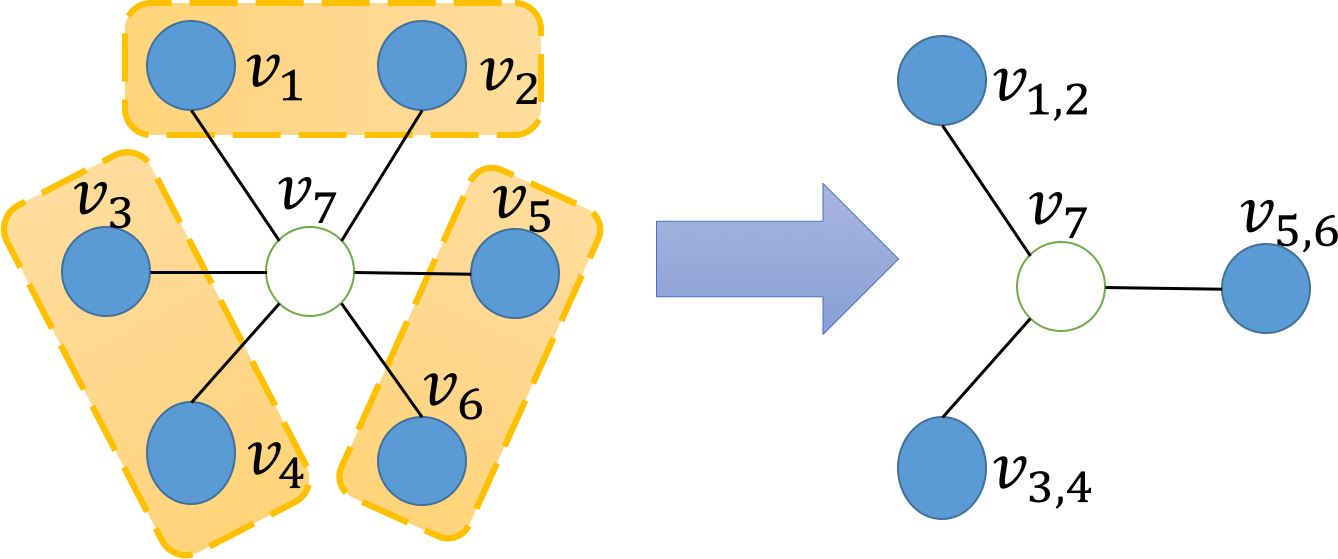}
\caption{Star Collapsing.}
\label{fig:star-collapsing}
\end{subfigure}
\caption{Illustration of graph coarsening algorithms.
\ref{fig:edge-collapsing}: Edge collapsing on a graph snippet.
\ref{fig:edge-collapsing-fail}: How edge collapsing fails to coalesce star-like structures.
\ref{fig:star-collapsing}: How star collapsing scheme coalesces the same graph snippet efficiently.}
\label{fig:viz-hierarchical-softmax}
\end{figure*}

\subsection{Algorithm: \texttt{\ouralgorithm}}

Our method for multi-level graph representation learning, \ouralgorithm, is presented in Algorithm \ref{alg:multilevel}.
It consists of three parts - graph coarsening, graph embedding, and representation refinement - which we detail below:

\begin{enumerate}
\item \emph{Graph Coarsening} (line 1):
Given a graph $G$, graph coarsening algorithms create a hierarchy of successively smaller graphs $G_0, G_1, \cdots, G_L$,
where $G_0 = G$.
The coarser (smaller) graphs preserve the global structure of the original graph, yet have significantly fewer nodes and edges.
Algorithms for generating this hierarchy of graphs will be discussed in detail below.

\item \emph{Graph Embedding on the Coarsest Graph} (line 2-3):
The graph embedding is obtained on the coarsest graph $G_L$ with the provided graph embedding algorithm.
As the size of $G_L$ is usually very small, it is much easier to get a high-quality graph representation.

\item \emph{Graph Representation Prolongation and Refinement} (line 4-7):
We prolong and refine the graph representation from the coarsest to the finest graph.
For each graph $G_i$, we prolong the graph representation of $G_{i + 1}$ as its initial embedding $\Phi_{G_i}^\prime$.
Then, the embedding algorithm $Embed()$ is applied to $(G_i, \Phi_{G_i}^\prime)$ to further refine $\Phi_{G_i}^\prime$, resulting in the refined embedding $\Phi_{G_i}$.
We discuss this step in the embedding prolongation section below.

\item \emph{Graph Embedding of the Original Graph} (line 8):
We return $\Phi_{G_0}$, which is the graph embedding of the original graph.
\end{enumerate}

\begin{algorithm}[t]
\begin{algorithmic}[1]
\Require
\Statex graph $G(V,E)$
\Statex arbitrary graph embedding algorithm $\Call{Embed}$
\Ensure matrix of vertex representations $\Phi \in \mathbb{R}^{|V| \times d}$
	\State $G_0, G_1, \cdots, G_L \leftarrow \Call{GraphCoarsening}{G}$
	\State Initialize $\Phi_{G_L}^\prime$ by assigning zeros
	\State $\Phi_{G_L} \leftarrow \Call{Embed}{G_L, \Phi_{G_L}^\prime}$
	\For{$i=L-1$ to $0$}
		\State $\Phi_{G_i}^\prime \leftarrow \Call{Prolongate}{\Phi_{G_{i+1}}, G_{i+1}, G_i}$
		\State $\Phi_{G_i} \leftarrow \Call{Embed}{G_i, \Phi_{G_i}^\prime}$
	\EndFor
\State \Return $\Phi_{G_0}$
\end{algorithmic}
\caption{\ouralgorithm($G, Embed()$)}
\label{alg:multilevel}
\end{algorithm}

We can easily see that this paradigm is algorithm independent, relying only on the provided functions $Embed()$.
Thus, with minimum effort, this paradigm can be incorporated into any existing graph representation learning methods, yielding a multilevel version of that method.

\subsection{Graph Coarsening}
\label{sec:coarsening}

\begin{algorithm}[t]
\begin{algorithmic}[1]
\Require graph $G(V,E)$
\Ensure Series of Coarsened Graphs $G_0, G_1, \cdots, G_L$
\State $L \leftarrow 0$
\State $G_0 \leftarrow G$
\While{$|V_{L}| \geq threshold$}
	\State $L \leftarrow L + 1$
	\State $G_{L} \leftarrow \Call{EdgeCollapse} {\Call{StarCollapse}{G} } $
\EndWhile
\State \Return $G_0, G_1, \cdots, G_{L}$
\label{alg:hybrid-collapsing}
\end{algorithmic}

\caption{GraphCoarsening($G$)}
\label{alg:graph-coarsening}
\end{algorithm}

In Algorithm \ref{alg:graph-coarsening}, we develop a hybrid graph coarsening scheme which preserves global graph structural information at different scales.
Its two key parts, namely edge collapsing and star collapsing, preserve \emph{first-order proximity} and \emph{second-order proximity} \cite{tang2015line} respectively.
First-order proximity is concerned with preserving the observed edges in the input graph,
while second-order proximity is based on the shared neighborhood structure of the nodes.

\textbf{Edge Collapsing.}
Edge collapsing \cite{hu2005efficient} is an efficient algorithm for preserving first-order proximity.
It selects $E^\prime \subseteq E$, such that no two edges in $E^\prime$ are incident to the same vertex.
Then, for each $(u_i, v_i) \in E^\prime$, it merges $(u_i, v_i)$ into a single node $w_i$, and merge the edges incident to $u_i$ and $v_i$.
The number of nodes in the coarser graph is therefore at least half of that in the original graph.
As illustrated in Figure \ref{fig:edge-collapsing},
the edge collapsing algorithm merges node pairs $(v_1, v_2)$ and $(v_3, v_4)$ into supernodes $v_{1, 2}$ and $v_{3, 4}$ respectively,
resulting in a coarser graph with 2 nodes and 1 edge.
The order of merging is arbitrary; we find different merging orders result in very similar node embeddings in practice.

\textbf{Star Collapsing.}
\label{sec:star-collapsing}
Real world graphs are often scale-free, which means they contain a large number of star-like structures.
A star consists of a popular central node (sometimes referred to as \textit{hubs}) connected to many peripheral nodes.
Although the edge collapsing algorithm is simple and efficient, it cannot sufficiently compress the star-like structures in a graph.
Consider the graph snippet in Figure \ref{fig:edge-collapsing-fail}, where the only central node $v_7$ connects to all the other nodes.
Assume the degree of the central node is $k$,
it is clear that the edge collapsing scheme can only compress this graph into a coarsened graph with $k - 1$ nodes.
Therefore when $k$ is large, the coarsening process could be arbitrarily slow, takes $O(k)$ steps instead of $O(\log k)$ steps.

One observation on the star structure is that
there are strong second-order similarities between the peripheral nodes since they share the same neighborhood.
This leads to our star collapsing scheme,
which merges nodes with the same neighbors into supernodes since they are similar to each other.
As shown in Figure \ref{fig:star-collapsing}, $(v_1, v_2)$, $(v_3, v_4)$ and $(v_5, v_6)$
are merged into supernodes as they share the same neighbors ($v_7$),
generating a coarsened graph with only $k / 2$ nodes.

\textbf{Hybrid Coarsening Scheme.}
By combining edge collapsing and star collapsing,
we present a hybrid scheme for graph coarsening in Algorithm \ref{alg:graph-coarsening}, which is adopted on all test graphs.
In each coarsening step, the hybrid coarsening scheme first decomposes the input graph with star collapsing,
then adopts the edge collapsing scheme to generate the coalesced graph.
We repeat this process until a small enough graph (with less than 100 vertices) is obtained.

\subsection{Embedding Prolongation}
\label{sec:prolongation}
After the graph representation for $G_{i+1}$ is learned, we prolong it into the initial representation for $G_{i}$.
We observe that each node $v \in G_{i+1}$ is either a member of the finer representation ($v \in G_{i}$), or the result of a merger, $(v_1, v_2, \cdots, v_k) \in G_{i}$.
In both cases, we can simply reuse the representation of the parent node $v \in G_i$ - the children are quickly separated by gradient updates.

\subsection{Complexity Analysis}
\label{sec:time_complexity}

In this section, we discuss the time complexity of \emph{\ourdw} and \emph{\ourline} and compare with the time complexity of \emph{DeepWalk} and \emph{LINE} respectively.
\emph{\ourntv} has the same time complexity as \emph{\ourdw}, thus it is not included in the discussion below.

\noindent \textbf{\ourdw}:
Given the number of random walks $\gamma$, walk length $t$, window size $w$ and representation size $d$,
the time complexity of \emph{DeepWalk} is dominated by the training time of the Skip-gram model, which is $\mathcal{O}(\gamma |V| t w (d + dlog|V|))$.
For \emph{\ourdw}, coarsening a graph with $|V|$ nodes produces a coarser graph with about $|V| / 2$ nodes.
The total number of nodes in all levels is approximately $|V|\sum_{i=0}^{log_2|V|}(\frac{1}{2})^{i}= 2|V|$.
Therefore, the time complexity of \emph{\ourdw} is $O(|V|)$ for copying binary tree and $\mathcal{O}(\gamma |V| t w (d + dlog|V|))$ for model training.
Thus, the overall time complexity of \emph{\ourdw} is also $\mathcal{O}(\gamma |V| t w (d + dlog|V|))$.

\noindent \textbf{\ourline}:
The time complexity of \emph{LINE} is linear to the number of edges in the graph and the number of iterations $r$ over edges, which is $\mathcal{O}(r |E|)$.
For \emph{\ourline}, coarsening a graph with $|E|$ nodes produces a coarsened graph with about $|E| / 2$ edges.
The total number edges in all levels is approximately $|E|\sum_{i=0}^{log_2|E|}(\frac{1}{2})^{i}= 2|E|$.
Thus, the time complexity of \emph{\ourline} is also $\mathcal{O}(r |E|)$.

\section{Experiment}
\label{Experiment}
In this section, we provide an overview of the datasets and methods used for experiments 
and evaluate the effectiveness of our method on challenging multi-label classification tasks in several real-life networks.
We further illustrate the scalability of our method and discuss its performance with regard to several important parameters.

\subsection{Datasets}
Table \ref{tab:table-graph-stats} gives an overview of the datasets used in our experiments.

\begin{table}
	\centering	
	{\footnotesize
    \begin{tabular}{l@{\quad}c c c c c}
    \toprule
	Name & DBLP & Blogcatalog & CiteSeer \\
    \midrule
    \# Vertices & 29,199 & 10,312 & 3,312 \\
    \# Edges & 133,664 & 333,983 & 4,732 \\
    \# Classes & 4 & 39 & 6 \\
    Task & Classification & Classification & Classification \\
    \bottomrule
    \end{tabular}
    }
    \caption{Statistics of the graphs used in our experiments.}
    \label{tab:table-graph-stats}
\end{table}

\begin{itemize}
\item{\emph{DBLP}} \cite{walklets} -- DBLP is a co-author graph of researchers in computer science.
The labels indicate the research areas a researcher publishes his work in.
The 4 research areas included in this dataset are DB, DM, IR, and ML.
\item{\emph{BlogCatalog}} \cite{tang2009relational} -- BlogCatalog is a network of social relationships between users on the BlogCatalog website. The labels represent the categories a blogger publishes in.
\item{\emph{CiteSeer}} \cite{sen:aimag08} -- CiteSeer is a citation network between publications in computer science.
The labels indicate the research areas a paper belongs to.
The papers are classified into 6 categories: Agents, AI, DB, IR, ML, and HCI.
\end{itemize}

\subsection{Baseline Methods}
We compare our model with the following graph embedding methods:

\begin{itemize}

\item \emph{DeepWalk} ---
\emph{DeepWalk} is a two-phase method for embedding graphs. Firstly, \emph{DeepWalk} generates random walks of fixed length from all the vertices of a graph. Then, the walks are treated as sentences in a language model and the Skip-Gram model for learning word embeddings is utilized to obtain graph embeddings. \emph{DeepWalk} uses hierarchical softmax for Skip-gram model optimization.

\item \emph{LINE} ---
\emph{LINE} is a method for embedding large-scale networks. The objective function of \emph{LINE} is designed for preserving both first-order and second-order proximities, and we use first-order \emph{LINE} for comparison. Skip-gram with negative sampling is used to solve the objective function.

\item \emph{Node2vec} ---
\emph{Node2vec} proposes an improvement to the random walk phase of \emph{DeepWalk}. By introducing the return parameter $p$ and the in-out parameter $q$, \emph{Node2vec} combines DFS-like and BFS-like neighborhood exploration.
\emph{Node2vec} also uses negative sampling for optimizing the Skip-gram model.

\end{itemize}
For each baseline method, we combine it with \emph{\ouralgorithm} and compare their performance.
\subsection{Parameter Settings}
Here we discuss the parameter settings for our models and baseline models.
Since \emph{DeepWalk}, \emph{LINE} and \emph{Node2vec} are all sampling based algorithms,
we always ensure that the total number of samples seen by the baseline algorithm
is the \textbf{same} as that of the corresponding \emph{\ouralgorithm} enhanced algorithm.

\textbf{DeepWalk.}
For \emph{DeepWalk} and \emph{\ourdw}, we need to set the following parameters:
the number of random walks $\gamma$, walk length $t$, window size $w$ for the Skip-gram model and representation size $d$.
In \emph{\ourdw}, the parameter setting is $\gamma = 40, t = 10, w = 10, d = 128$.
For \emph{DeepWalk}, all the parameters except $\gamma$ are the same as in \emph{\ourdw}.
Specifically, to ensure a fair comparison, we increase the value of $\gamma$ for \emph{DeepWalk}.
This gives \emph{DeepWalk} a larger training dataset (as large as all of the levels of \emph{\ourdw} combined).
We note that failure to increase $\gamma$ in this way resulted in substantially worse \emph{DeepWalk} (and \emph{Node2vec}) models.

\textbf{LINE.}
For \emph{\ourline}, we run $50$ iterations on all graph edges on all coarsening levels.
For \emph{LINE}, we increase the number of iterations over graph edges accordingly, so that the amount of training data for both models remain the same.
The representation size $d$ is set to $64$ for both \emph{LINE} and \emph{\ourline}.

\textbf{Node2vec.}
For \emph{\ourntv}, the parameter setting is $\gamma = 40, t = 10, w = 10, d = 128$.
Similar to \emph{DeepWalk}, we increase the value of $\gamma$ in \emph{Node2vec} to ensure a fair comparison.
Both in-out and return hyperparameters are set to 1.0.
For all models, the initial learning rate and final learning rate are set to 0.025 and 0.001 respectively.

\subsection{Graph Coarsening}
\begin{figure}[t]
    \centering	
    \begin{subfigure}[b]{.32\linewidth} 
        \includegraphics[width=\linewidth]{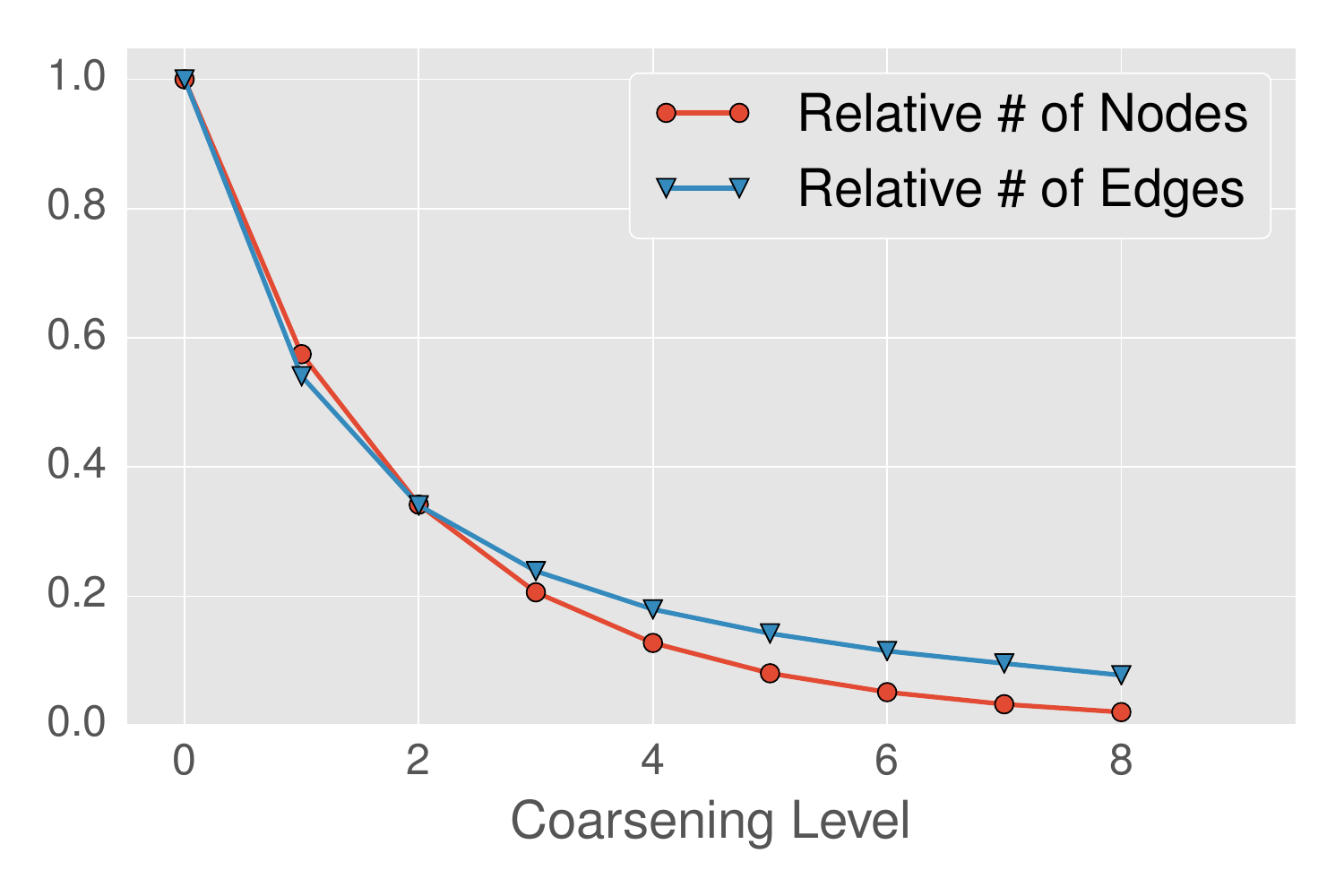}
        \caption{DBLP}        
	\end{subfigure}
    \begin{subfigure}[b]{.32\linewidth}
        \includegraphics[width=\linewidth]{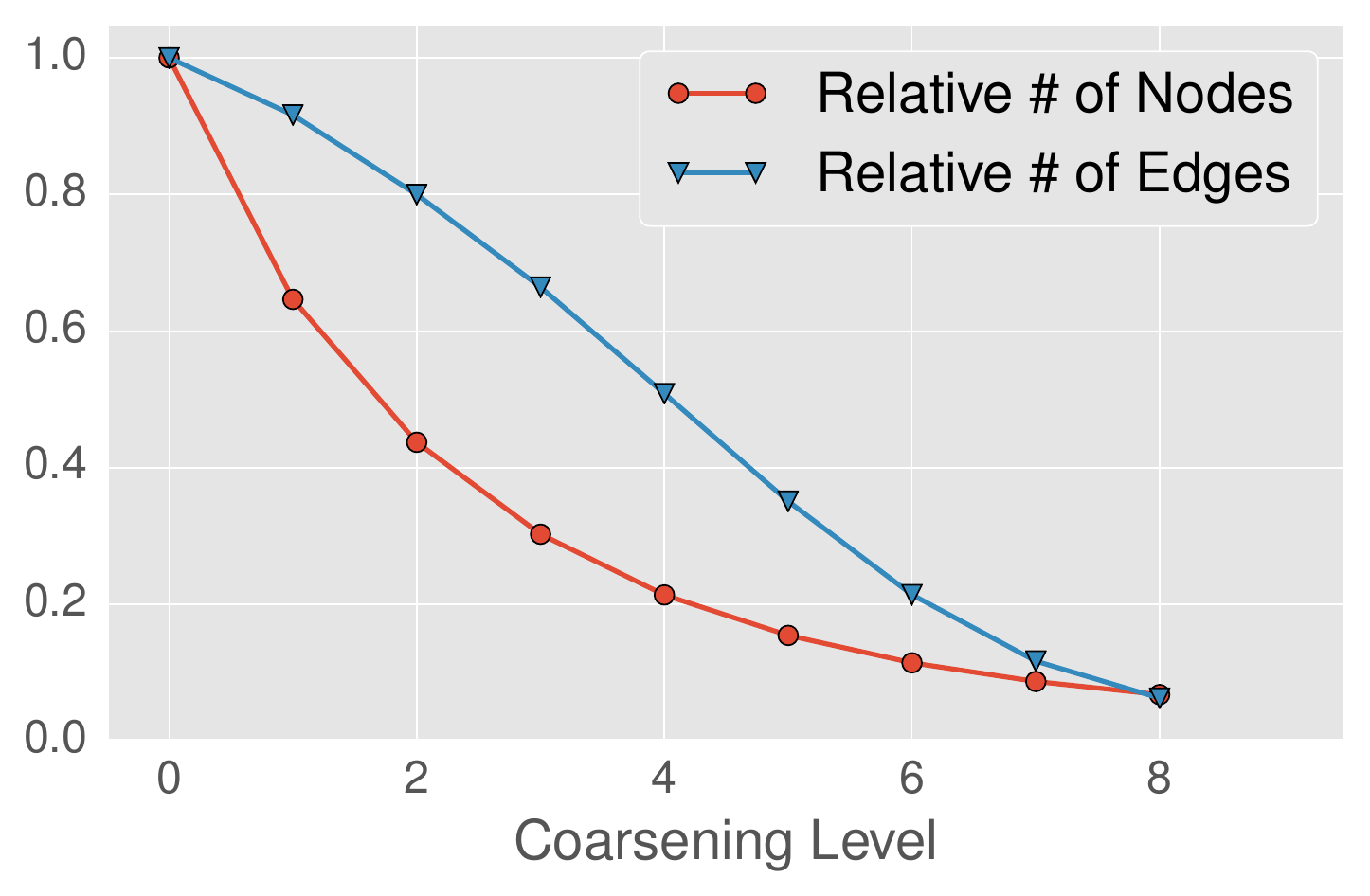}
        \caption{BlogCatalog}        
	\end{subfigure}
    \begin{subfigure}[b]{.32\linewidth} 
        \includegraphics[width=\linewidth]{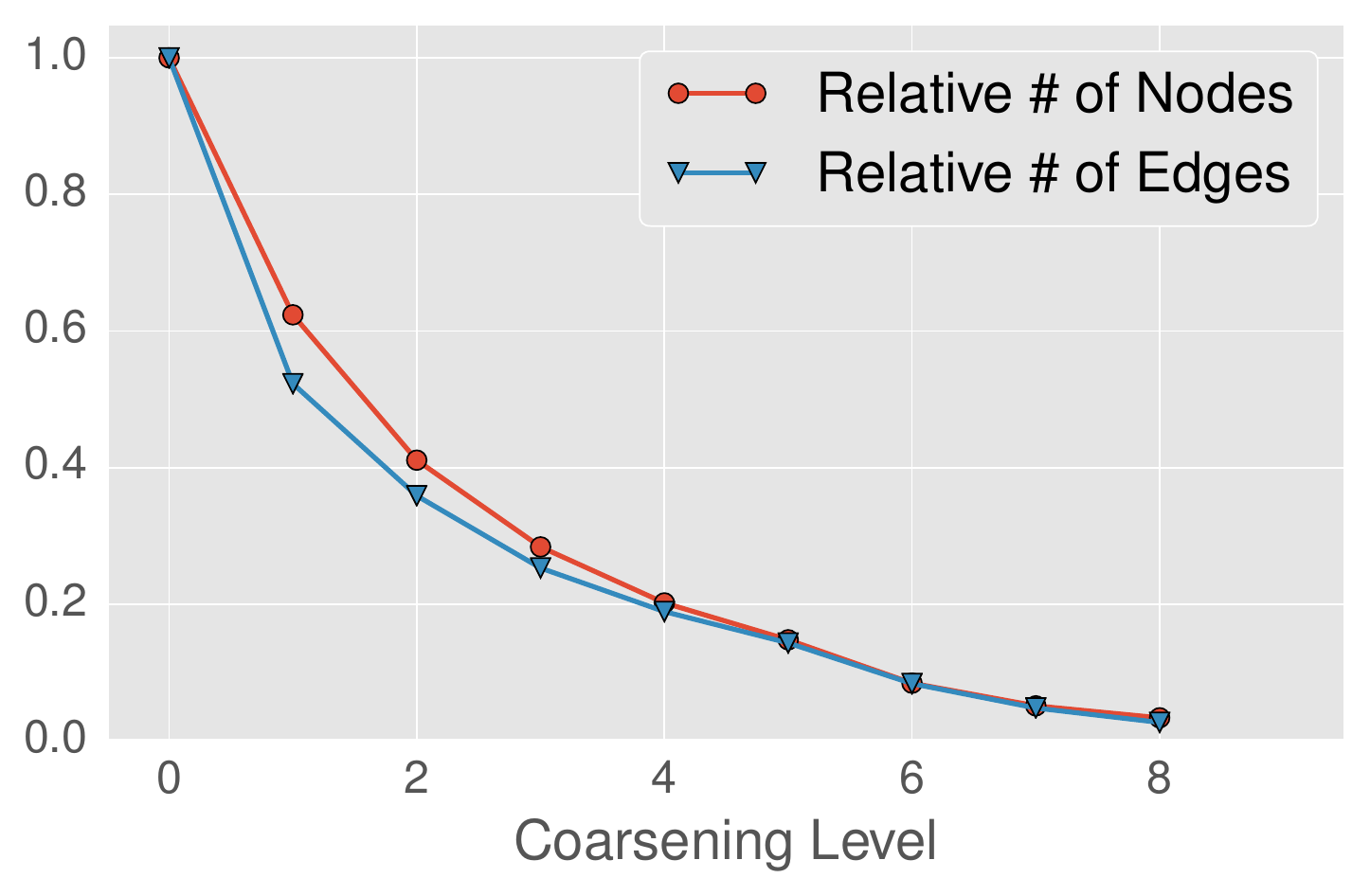}
        \caption{CiteSeer}        
	\end{subfigure}
    \caption{The ratio of nodes/edges of the coarsened graphs to that of the original test graphs.
	For disconnected graphs, the graph coarsening result on the largest connected component is shown.}
    \label{fig:coarsening_stats}
\end{figure}

Figure \ref{fig:coarsening_stats} demonstrates the effect of our hybrid coarsening method on all test graphs. 
The first step of graph coarsening for each graph eliminates about half the nodes,
but the number of edges only reduce by about 10\% for \emph{BlogCatalog}.
This illustrates the difficulty of coarsening real-world graphs.
However, as the graph coarsening process continues, the scale of all graphs drastically decrease.
At level 8, all graphs have less than 10\% nodes and edges left.

\subsection{Visualization}
To show the intuition of the \emph{\ouralgorithm} paradigm, we set $d = 2$, and visualize the graph representation generated by \emph{\ourline} at each level.

\begin{figure}
\centering
\begin{subfigure}[b]{.20\linewidth}
\includegraphics[width=\linewidth]{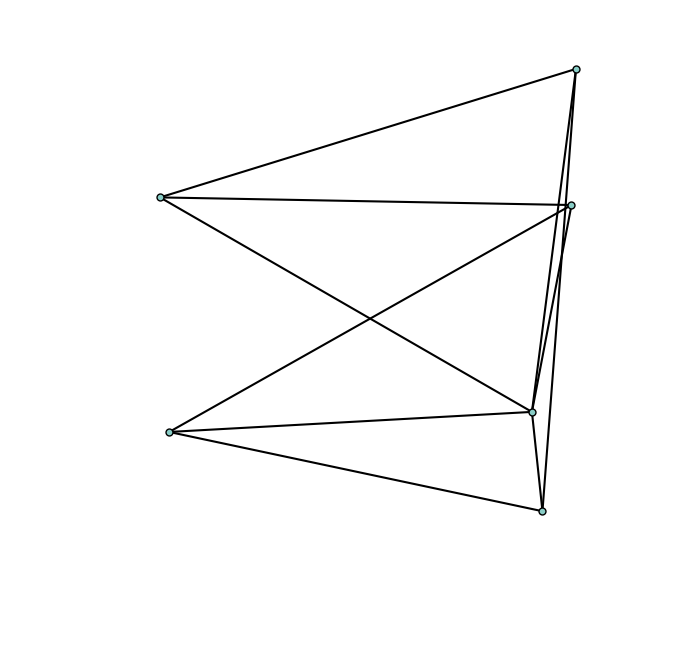}
\caption{Level 7}\label{fig:poisson_2d_gc_8}
\end{subfigure}
\begin{subfigure}[b]{.20\linewidth}
\includegraphics[width=\linewidth]{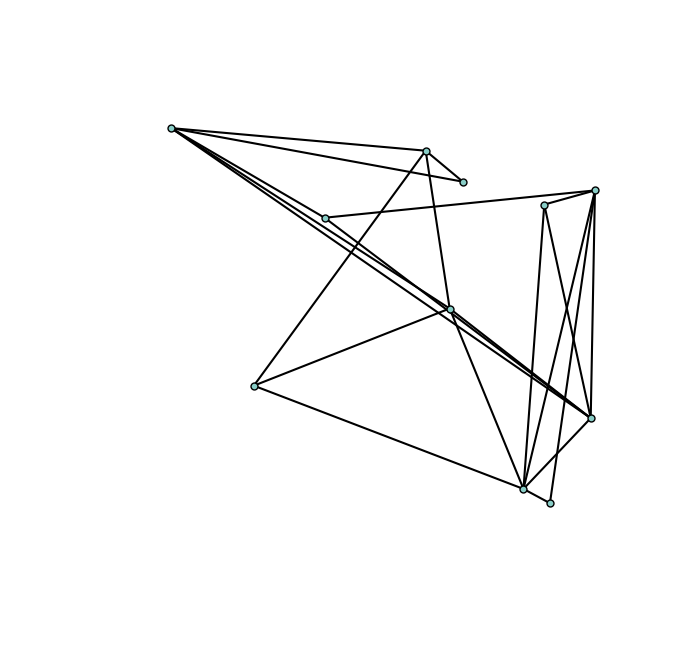}
\caption{Level 6}\label{fig:poisson_2d_gc_7}
\end{subfigure}
\begin{subfigure}[b]{.20\linewidth}
\includegraphics[width=\linewidth]{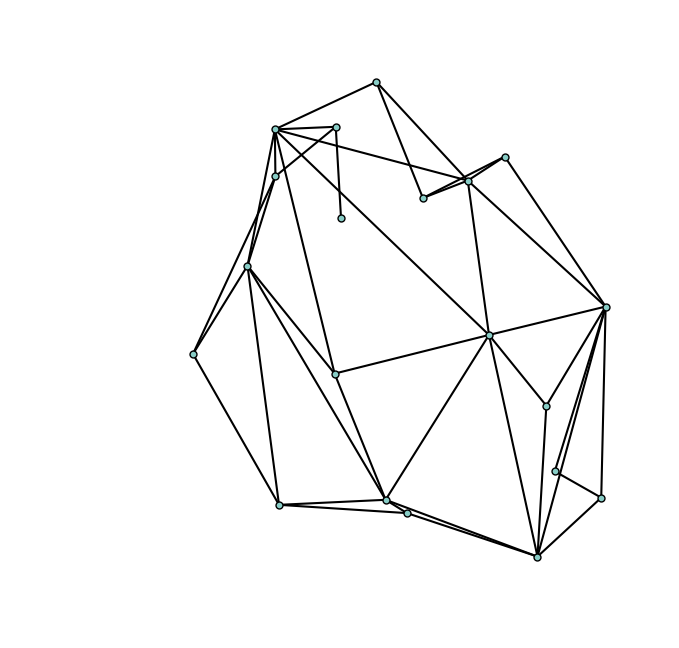}
\caption{Level 5}\label{fig:poisson_2d_gc_6}
\end{subfigure}

\begin{subfigure}[b]{.20\linewidth}
\includegraphics[width=\linewidth]{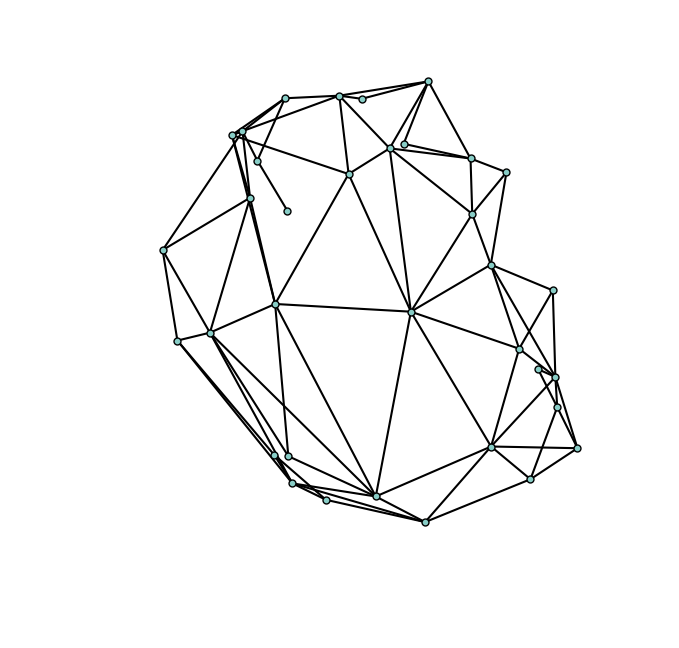}
\caption{Level 4}\label{fig:poisson_2d_gc_5}
\end{subfigure}
\begin{subfigure}[b]{.20\linewidth}
\includegraphics[width=\linewidth]{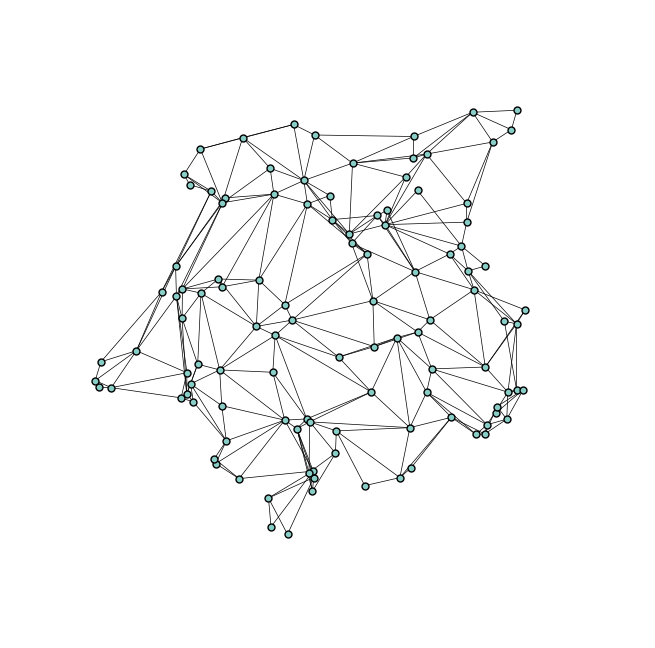}
\caption{Level 3}\label{fig:poisson_2d_gc_4}
\end{subfigure}
\begin{subfigure}[b]{.20\linewidth}
\includegraphics[width=\linewidth]{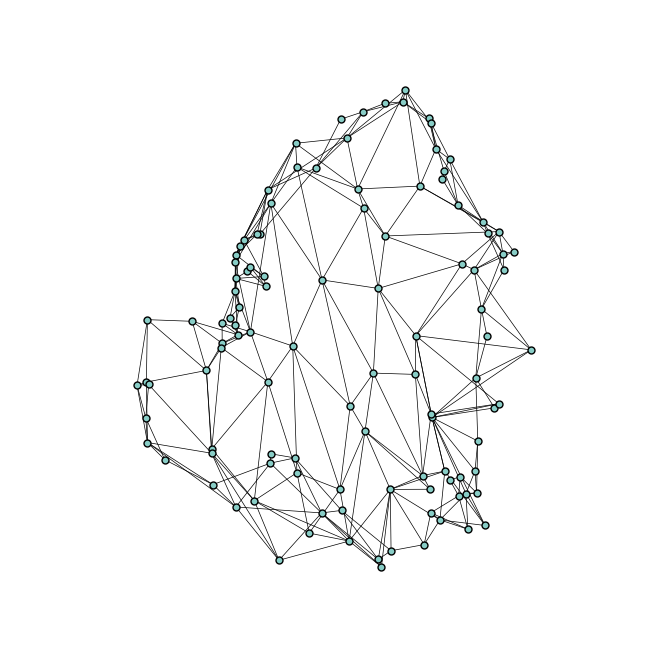}
\caption{Level 2}\label{fig:poisson_2d_gc_3}
\end{subfigure}

\begin{subfigure}[b]{.20\linewidth}
\includegraphics[width=\linewidth]{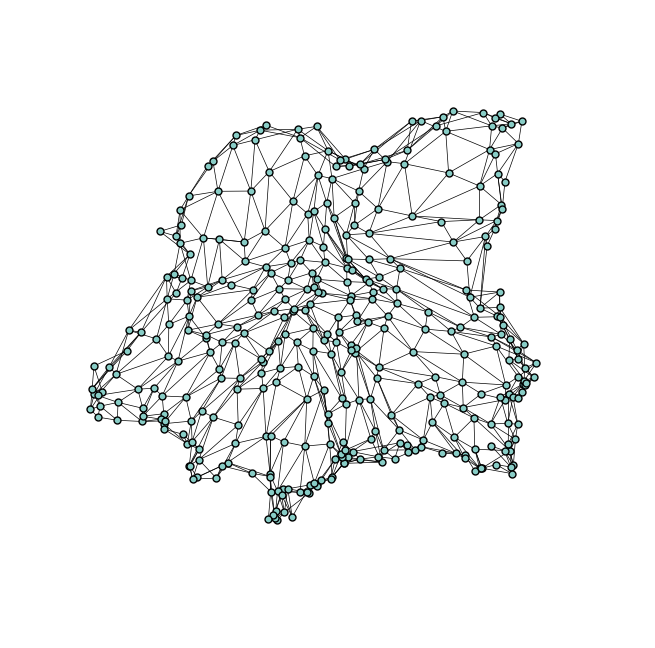}
\caption{Level 1}\label{fig:poisson_2d_gc_2}
\end{subfigure}
\begin{subfigure}[b]{.20\linewidth}
\includegraphics[width=\linewidth]{figures/poisson_2d_gc_1.png}
\caption{Level 0}\label{fig:poisson_2d_gc_1}
\end{subfigure}
\begin{subfigure}[b]{.20\linewidth}
\includegraphics[width=\linewidth]{figures/poisson_2d_gd.png}
\caption{Input}\label{fig:poisson_2d_sfdp_dup}
\end{subfigure}

\caption{Two-dimensional embeddings generated with \emph{\ourline} on different coarsening levels on \emph{Poisson 2D}.
Level 7 denotes the smallest graph, while level 0 denotes the original graph. 
The last subfigure is the graph layout generated by a force-direct graph drawing algorithm.}
\label{fig:poisson_2d_line_gc_detailed}
\end{figure}

Figure \ref{fig:poisson_2d_line_gc_detailed} shows the level-wise 2D graph embeddings obtained with \emph{\ourline} on \emph{Poisson 2D}.
The graph layout of level 5 (which has only 21 nodes) already highly resembles the layout of the original graph.
The graph layout on each subsequent level is initialized with the prolongation of the previous graph layout, thus the global structure is kept.

\subsection{Multi-label Classification}
We evaluate our method using the same experimental procedure in \cite{perozzi2014deepwalk}.
Firstly, we obtain the graph embeddings of the input graph.
Then, a portion ($T_R$) of nodes along with their labels are randomly sampled from the graph as training data,
and the task is to predict the labels for the remaining nodes.
We train a one-vs-rest logistic regression model with L2 regularization on the graph embeddings for prediction.
The logistic regression model is implemented by LibLinear \cite{fan2008liblinear}.
To ensure the reliability of our experiment, the above process is repeated for 10 times, and the average Macro $F_1$ score is reported.
The other evaluation metrics such as Micro $F_1$ score and accuracy follow the same trend as Macro $F_1$ score,
thus are not shown.
\begin{table}[t!]
\centering
\begin{tabular}{clll}
\toprule
\textbf{Algorithm} & \multicolumn{3}{c}{\textbf{Dataset}}           						\\ 
                                 & DBLP                   & BlogCatalog    & CiteSeer       \\ \midrule
\emph{DeepWalk}                  & 57.29                  & 24.88          & 42.72          \\
\emph{\ourdw}                    & {$\mathbf{61.76^*}$}         & $\mathbf{25.90^*}$ & $\mathbf{44.78^*}$ \\
\emph{Gain of \ouralgorithm [\%]}     & \textbf{7.8}           & \textbf{4.0}   & \textbf{4.8}   \\ \midrule
\emph{LINE}                      & 57.76                  & 22.43          & 37.11          \\
\emph{\ourline}                  & $\mathbf{59.51^*}$         & $\mathbf{23.47^*}$ & $\mathbf{42.95^*}$ \\
\emph{Gain of \ouralgorithm [\%]}     & \textbf{3.0}           & \textbf{4.6}   & \textbf{13.6}  \\ \midrule
\emph{Node2vec}                  & 62.64         		  & 23.55          & 44.84          \\
\emph{\ourntv}                   & \textbf{62.80}         & $\mathbf{24.66^*}$ & $\mathbf{46.08^*}$ \\
\emph{Gain of \ouralgorithm [\%]}     & \textbf{0.3}           & \textbf{4.7}   & \textbf{2.8}   \\
\bottomrule
\end{tabular}
\caption{Macro $F_1$ scores and performance gain of \emph{\ouralgorithm} on \emph{DBLP}, \emph{BlogCatalog}, and \emph{CiteSeer} in percentage.
* indicates statistically superior performance to the corresponding baseline method at level of 0.001 using a standard paired t-test.
Our method improves \textbf{all} existing neural embedding techniques.}
\label{tab:classification_summary}
\end{table}

\begin{figure*}[t]
    \centering

    \begin{subfigure}[b]{.3\linewidth}
        \includegraphics[width=\linewidth]{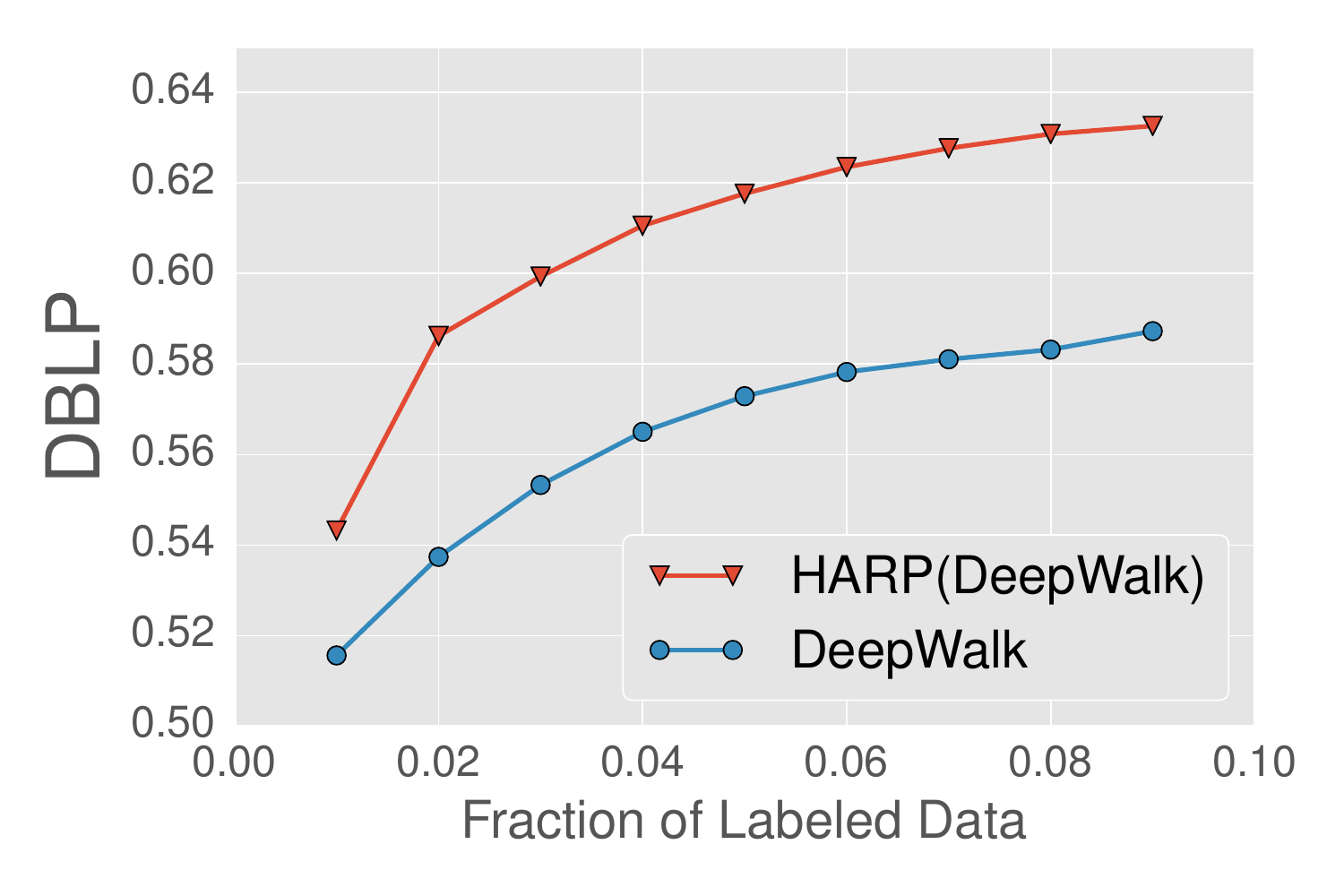}
    \end{subfigure}
    \begin{subfigure}[b]{.3\linewidth}
        \includegraphics[width=\linewidth]{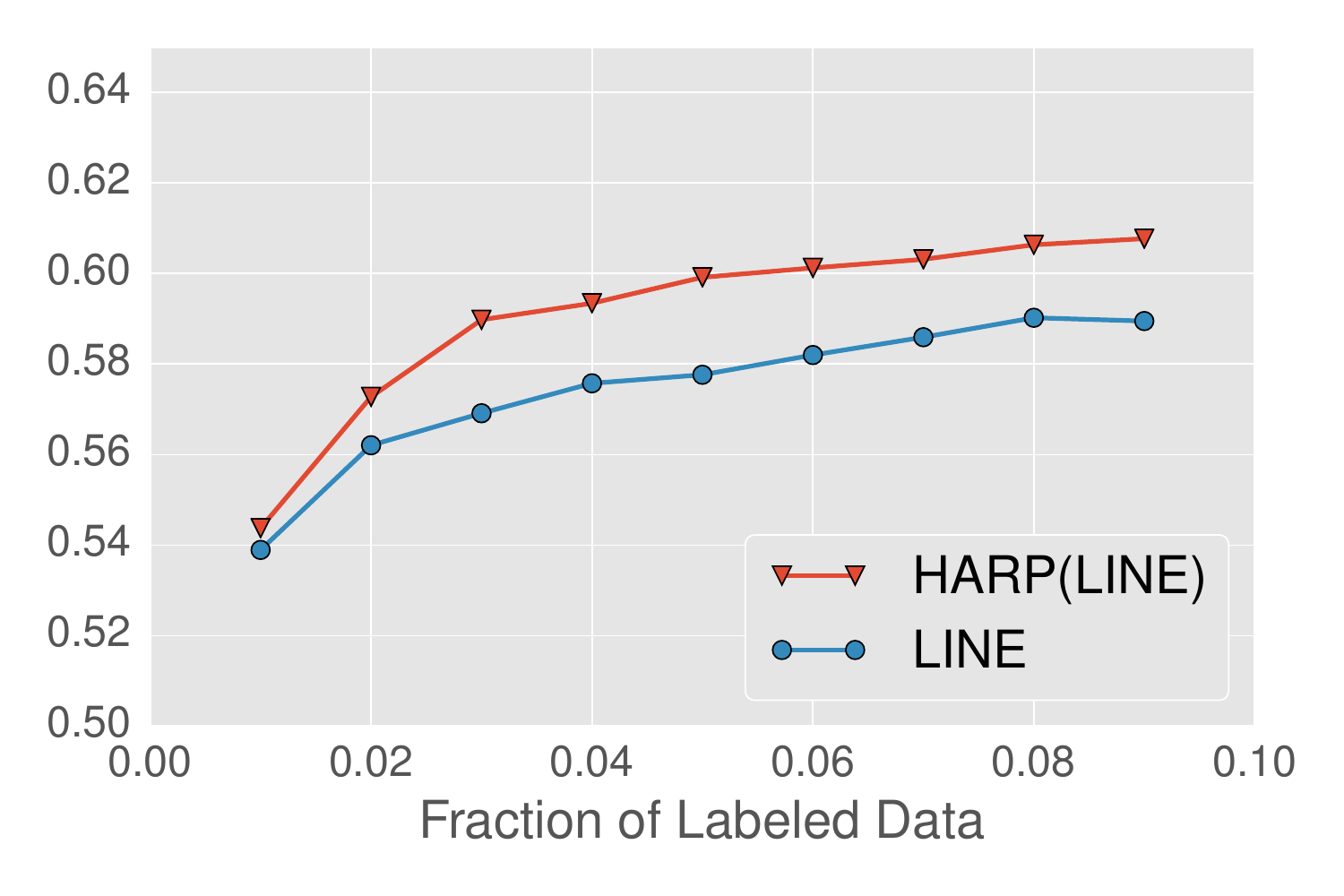}
    \end{subfigure}
    \begin{subfigure}[b]{.3\linewidth}
        \includegraphics[width=\linewidth]{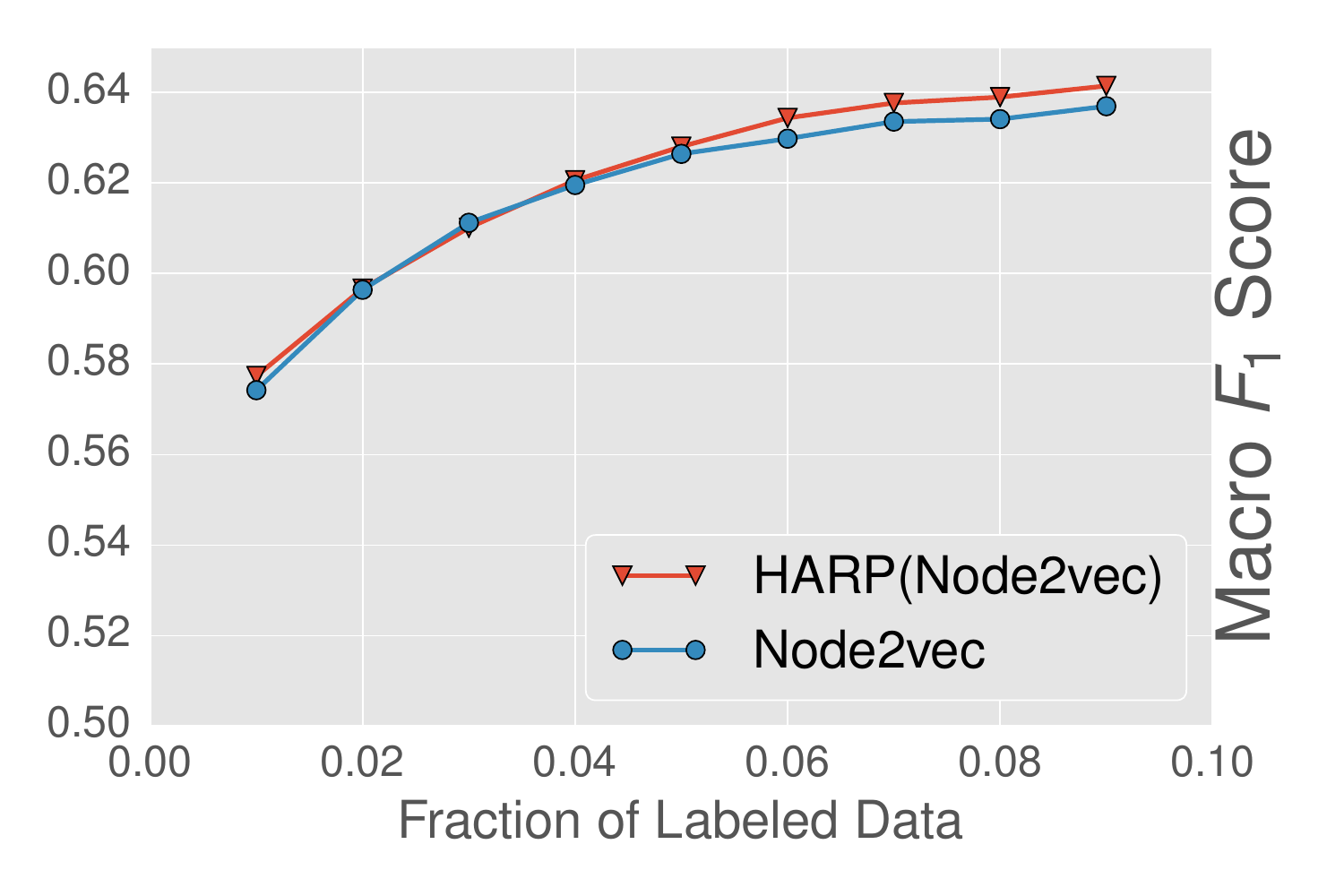}
	\end{subfigure}

    \begin{subfigure}[b]{.3\linewidth}
        \includegraphics[width=\linewidth]{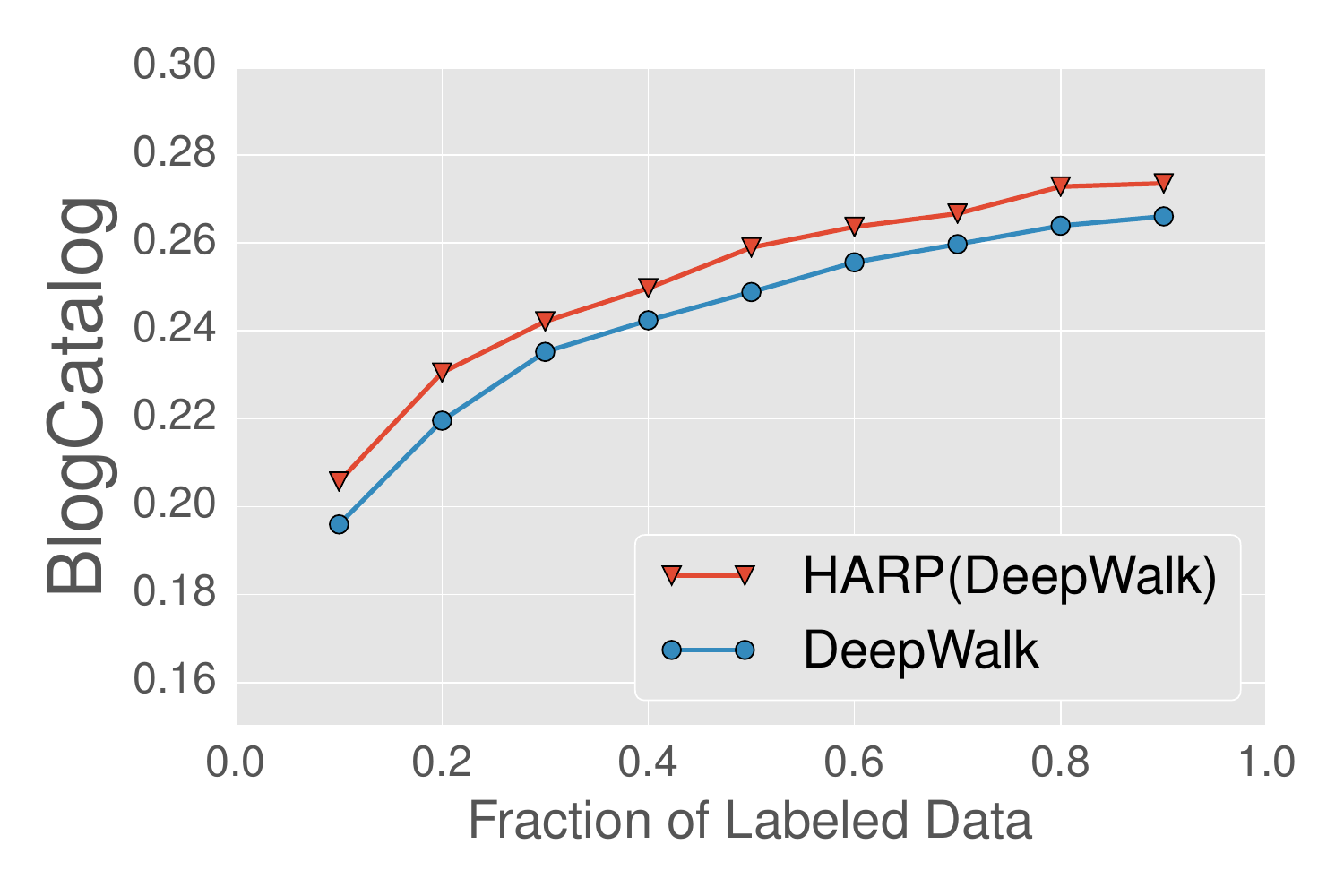}
    \end{subfigure}
    \begin{subfigure}[b]{.3\linewidth}
        \includegraphics[width=\linewidth]{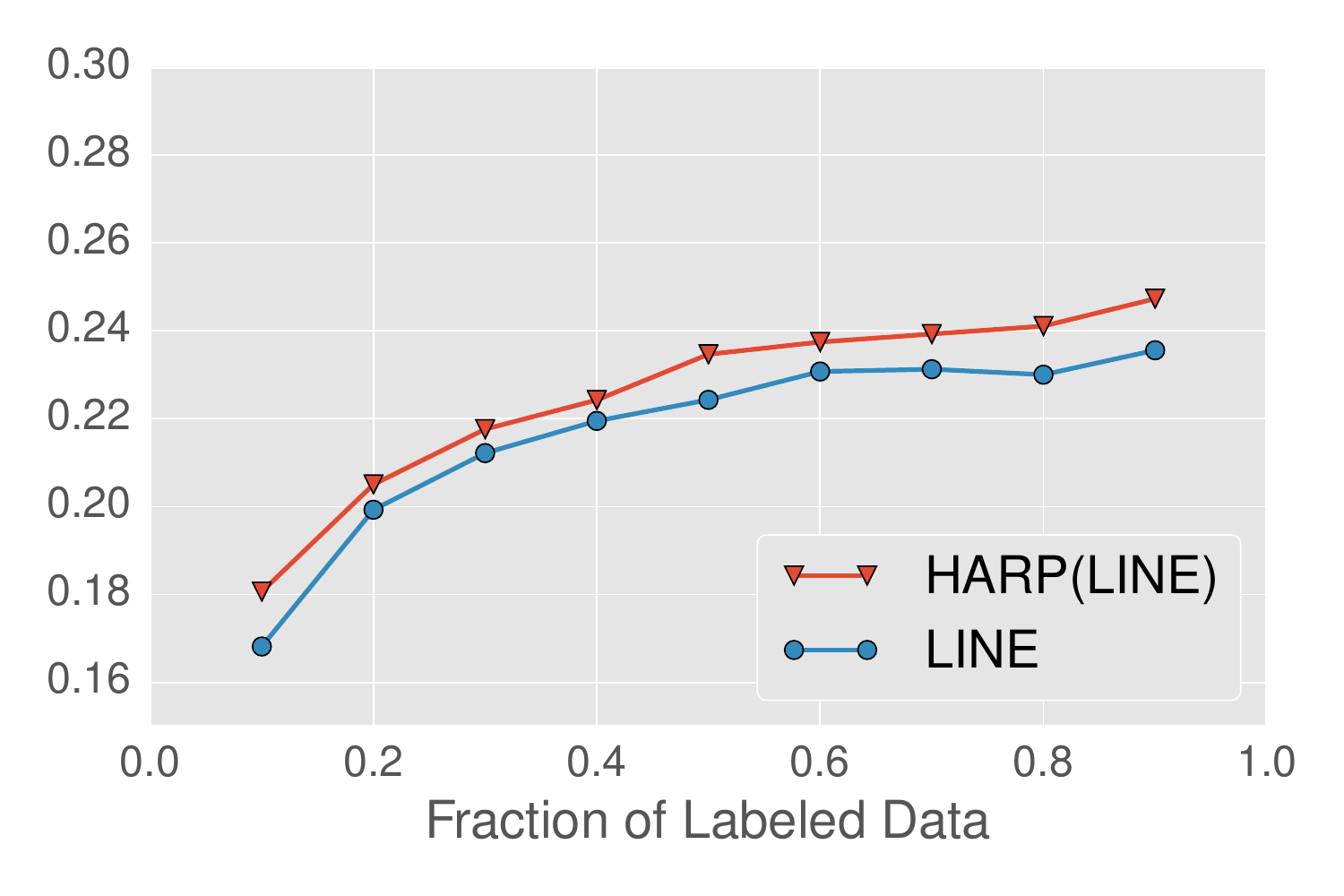}
    \end{subfigure}
    \begin{subfigure}[b]{.3\linewidth}
        \includegraphics[width=\linewidth]{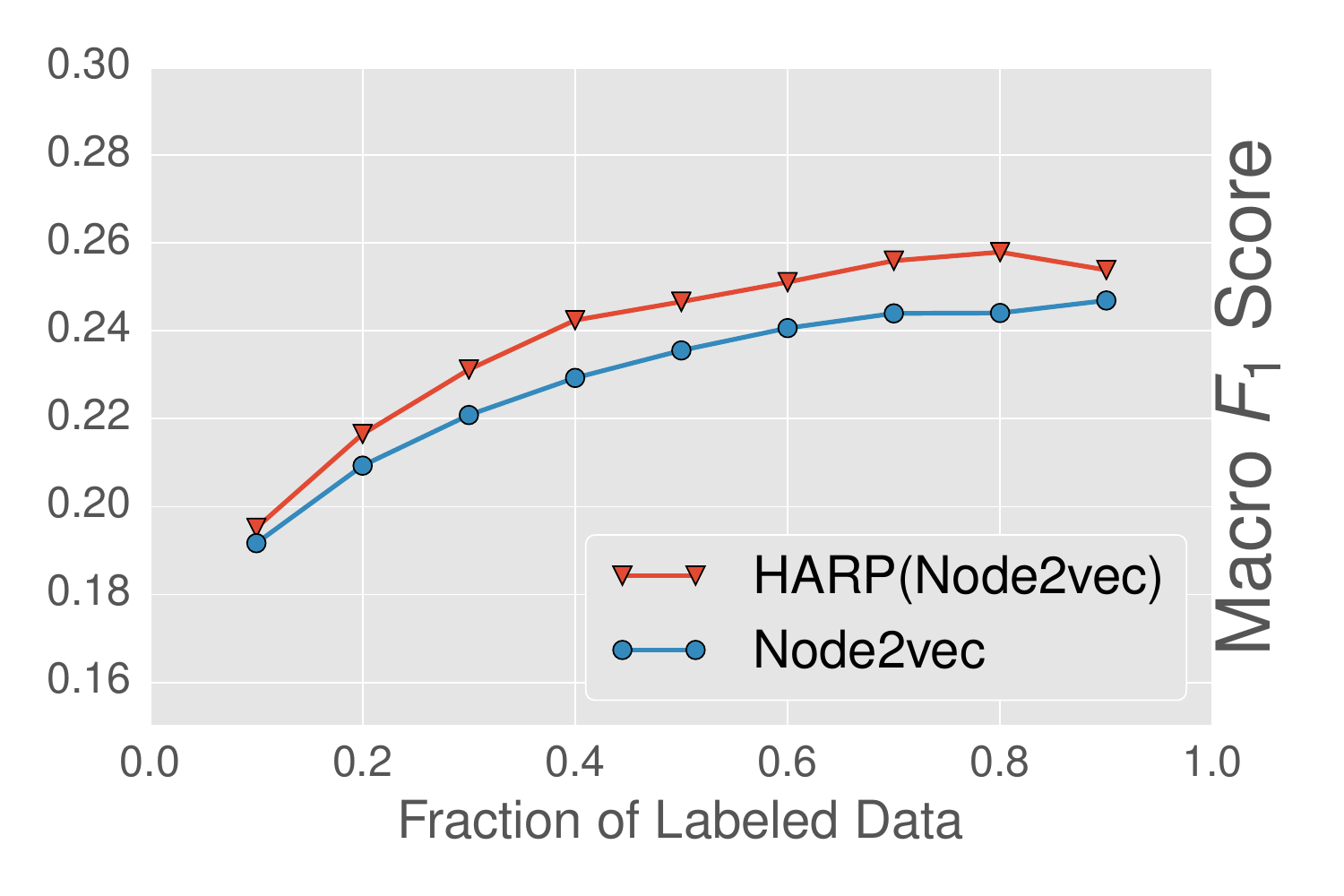}
	\end{subfigure}

    \begin{subfigure}[b]{.3\linewidth}
        \includegraphics[width=\linewidth]{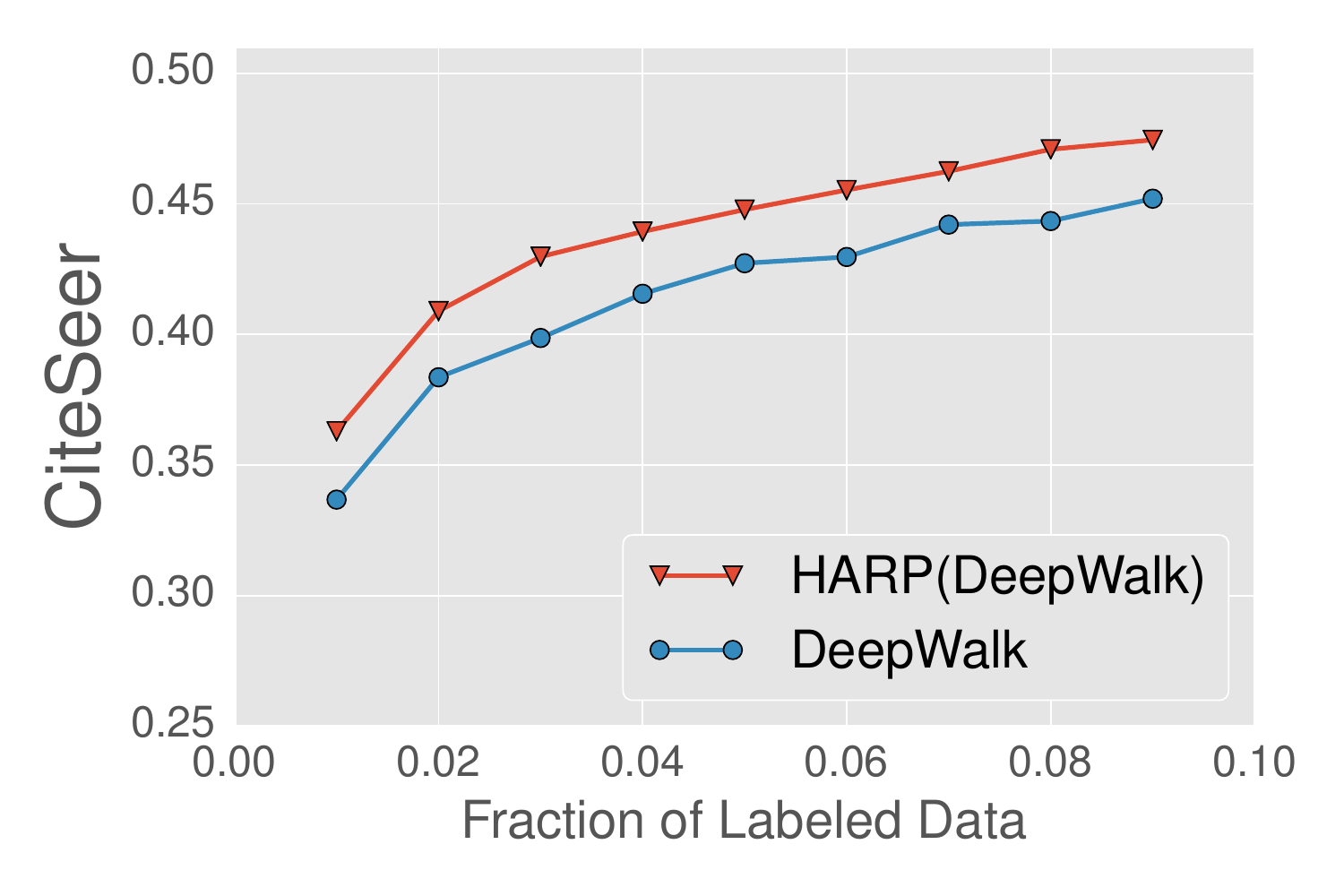}
    \end{subfigure}
    \begin{subfigure}[b]{.3\linewidth}
        \includegraphics[width=\linewidth]{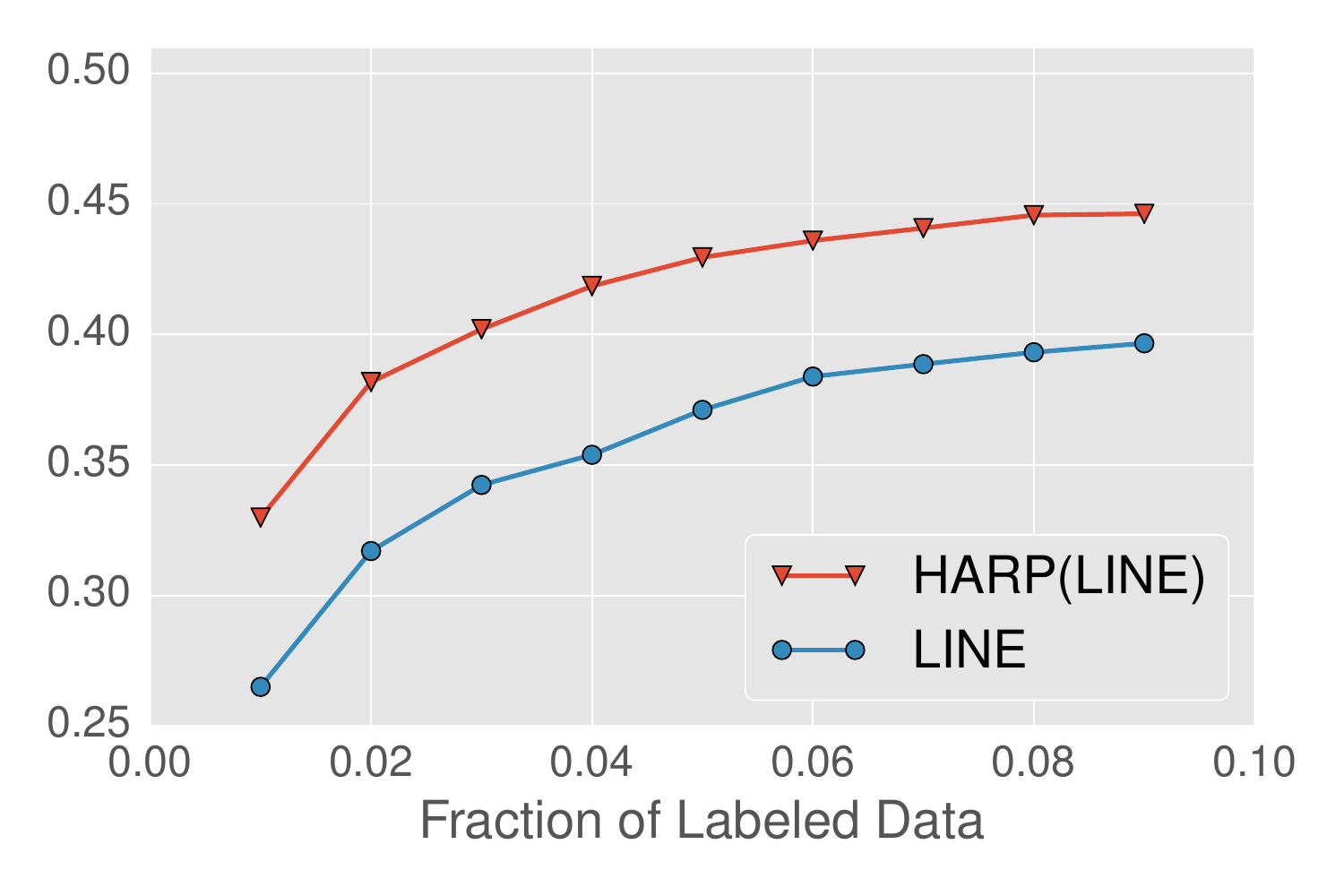}
    \end{subfigure}
    \begin{subfigure}[b]{.3\linewidth}
        \includegraphics[width=\linewidth]{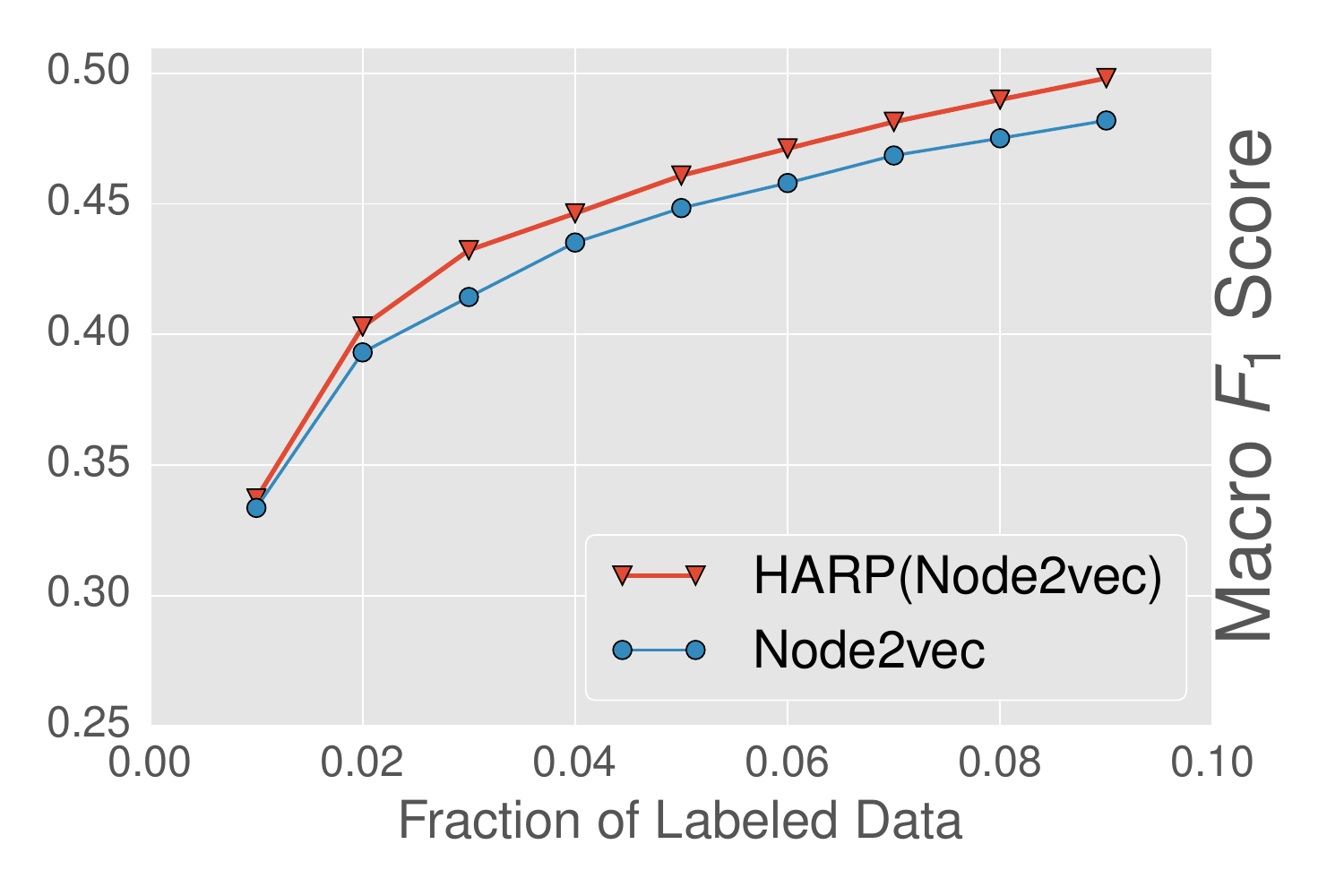}
	\end{subfigure}

    \caption{Detailed multi-label classification result on \emph{DBLP}, \emph{BlogCatalog}, and \emph{CiteSeer}.}
    \label{fig:classification_details}

\end{figure*}
Table \ref{tab:classification_summary} reports the Macro $F_1$ scores achieved on \emph{DBLP}, \emph{BlogCatalog}, and \emph{CiteSeer} with 5\%, 50\%, and 5\% labeled nodes respectively.
The number of class labels of \emph{BlogCatalog} is about 10 times that of the other two graphs,
thus we use a larger portion of labeled nodes.
We can see that our method improves all existing neural embedding techniques on all test graphs.
In \emph{DBLP}, the improvements introduced by \emph{\ourdw}, \emph{\ourline} and \emph{\ourntv}
are 7.8\%, 3.0\% and 0.3\% respectively.
Given the scale-free nature of \emph{BlogCatalog}, graph coarsening is much harder due to a large amount of star-like structures in it.
Still, \emph{\ourdw}, \emph{\ourline} and \emph{\ourntv} achieve gains of
4.0\%, 4.6\% and 4.7\% over the corresponding baseline methods respectively.
For \emph{CiteSeer}, the performance improvement is also striking: \emph{\ourdw}, \emph{\ourline} and \emph{\ourntv}
outperforms the baseline methods by 4.8\%, 13.6\%, and 2.8\%.

To have a detailed comparison between \ouralgorithm\ and the baseline methods, we vary the portion of labeled nodes for classification,
and present the macro $F_1$ scores in Figure \ref{fig:classification_details}.
We can observe that \emph{\ourdw}, \emph{\ourline} and \emph{\ourntv}
consistently perform better than the corresponding baseline methods.

\textbf{DBLP.}
For \emph{DBLP}, the relative gain of \emph{\ourdw} is over 9\% with 4\% labeled data.
With only 2\% labeled data, \emph{\ourdw} achieves higher macro $F_1$ score than
\emph{DeepWalk} with 8\% label data.
\emph{\ourline} also consistently outperforms \emph{LINE} given any amount of training data,
with macro $F_1$ score gain between 1\% and 3\%.
\emph{\ourntv} and \emph{Node2vec} have comparable performance with less than 5\% labeled data,
but as the ratio of labeled data increases,
\emph{\ourntv} eventually distances itself to a 0.7\% improvement over \emph{Node2vec}.
We can also see that \emph{Node2vec} generally has better performance when compared to \emph{DeepWalk},
and the same holds for \emph{\ourntv} and \emph{\ourdw}.
The difference in optimization method for Skip-gram (negative sampling for \emph{Node2vec} and hierarchical softmax for \emph{DeepWalk})
may account for this difference.

\textbf{BlogCatalog.}
As a scale-free network with complex structure, \emph{BlogCatalog} is challenging for graph coarsening.
Still, by considering both first-order proximity and second-order proximity,
our hybrid coarsening algorithm generates an appropriate hierarchy of coarsened graphs.
With the same amount of training data, \emph{\ourdw} always leads by at least 3.0\%.
For \emph{\ourline}, it achieves a relative gain of 4.8\% with 80\% labeled data.
For \emph{\ourntv}, its gain over \emph{Node2vec} reaches 4.7\% given 50\% labeled nodes.

\textbf{Citeseer.}
For \emph{CiteSeer}, the lead of \emph{\ourdw} on Macro $F_1$ score varies between 5.7\% and 7.8\%.
For \emph{\ourline}, its improvement over \emph{LINE} with 4\% labeled data is an impressive 24.4\%.
\emph{\ourntv} also performs better than \emph{Node2vec} on any ratio of labeled nodes.

\subsection{Scalability}

\begin{figure}[t]
	\hspace{-0.1in}
    \centering
    \begin{subfigure}[b]{.49\linewidth}
		\includegraphics[width=\linewidth]{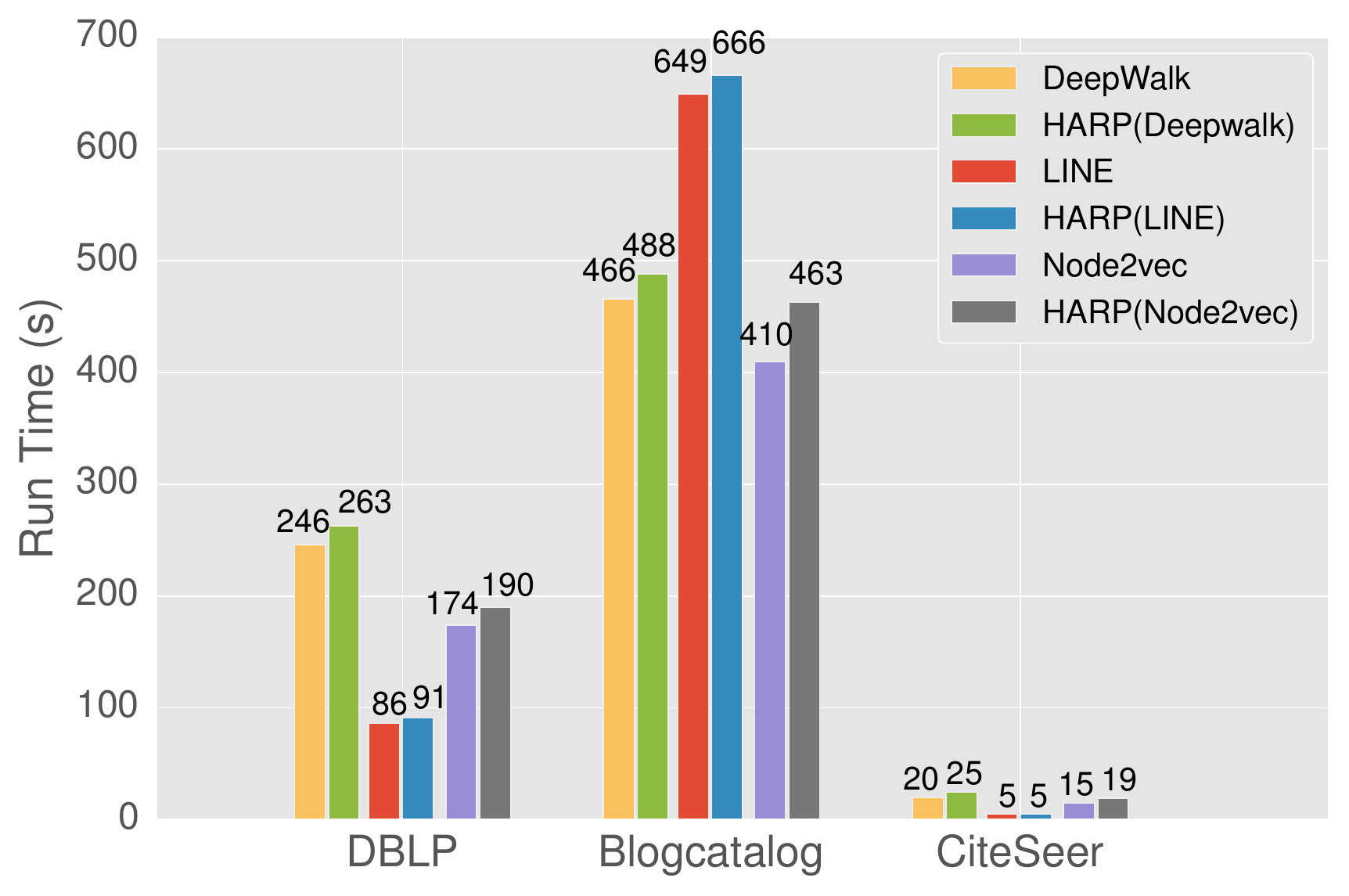}
		\caption{Test graphs.}
	    \label{fig:run_time_all}
    \end{subfigure}    
    \begin{subfigure}[b]{.49\linewidth}
		\includegraphics[width=\linewidth]{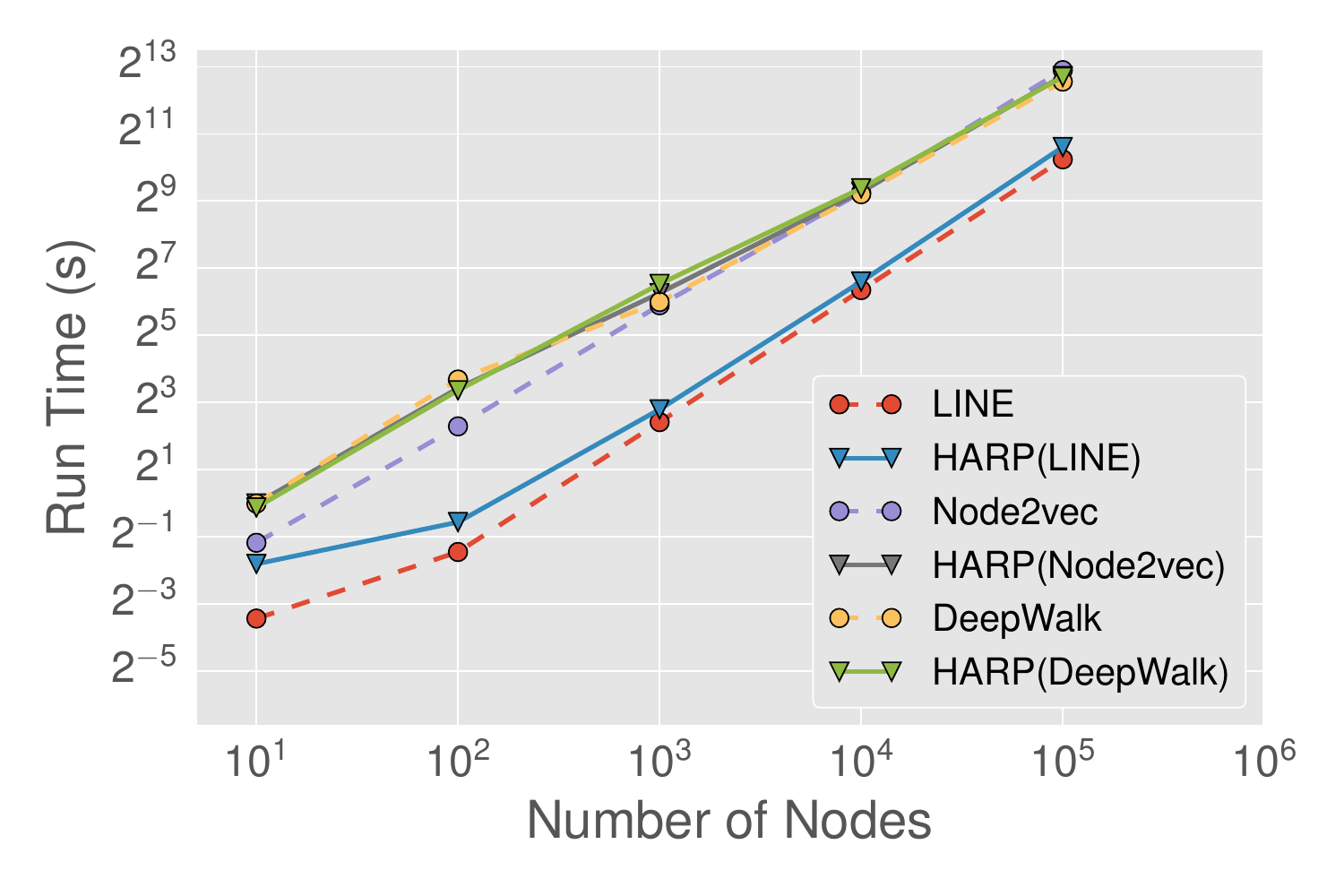}
		\caption{Erdos-Renyi graphs.}
	    \label{fig:run_time_er}
    \end{subfigure}

\caption{Runtime analysis.}
\label{fig:run_time}
\end{figure}

We already shown that introducing \emph{\ouralgorithm} does not affect the time complexity of the underlying graph embedding algorithms.
Here, we compare the actual run time of \emph{\ouralgorithm} enhanced embedding algorithms with the corresponding baseline methods on all test graphs.
All models run on a single machine with 128GB memory, 24 CPU cores at 2.0GHZ with 20 threads.
As shown in Figure \ref{fig:run_time_all}, applying \emph{\ouralgorithm} typically only introduces an overhead of less than 10\% total running time.
The time spent on sampling and training the Skip-gram model dominates the overall running time.

Additionally, we learn graph embeddings on Erdos-Renyi graphs with node count ranging from 100 to 100,000 and constant average degree of 10.
In Figure \ref{fig:run_time_er}, we can observe that the running time of \emph{\ouralgorithm} increases linearly with the number of nodes in the graph.
Also, when compared to the corresponding baseline method,
the overhead introduces by the graph coarsening and prolongation process in \ouralgorithm\ is negligible, especially on large-scale graphs.

\section{Related Work}
\label{Related_Work}
The related work is in the areas of graph representation learning and graph drawing, which we briefly describe here.

\noindent
\textbf{Graph Representation Learning}.
Most early methods treated representation learning as performing dimension reduction on the Laplacian and adjacency matrices \cite{belkin2001laplacian,cox2000multidimensional,tenenbaum2000global}. 
These methods work well on small graphs, but the time complexity of these algorithms is too high for the large-scale graphs commonly encountered today.

Recently, neural network-based methods have been proposed for constructing node representation in large-scale graphs. Deepwalk \cite{perozzi2014deepwalk} presents a two-phase algorithm for graph representation learning. In the first phase, Deepwalk samples sequences of neighboring nodes of each node by random walking on the graph. Then, the node representation is learned by training a Skip-gram model \cite{mikolov2013distributed} on the random walks. 
A number of methods have been proposed which extend this idea.
First, several methods use different strategies for sampling neighboring nodes.
LINE \cite{tang2015line} learns graph embeddings which preserve both the first-order and second-order proximities in a graph. 
Walklets \cite{walklets} captures multiscale node representation on graphs by sampling edges from higher powers of the graph adjacency matrix. 
Node2vec \cite{node2vec-kdd2016} combines DFS-like and BFS-like exploration within the random walk framework.
Second, matrix factorization methods and deep neural networks have also been proposed \cite{cao2015grarep,ouasymmetric,wangstructural,abu2017learning} as alternatives to the Skip-gram model for learning the latent representations.

Although these methods are highly scalable, they all rely on optimizing a non-convex objective function.
With no prior knowledge of the graph, the latent representations are usually initialized with random numbers or zero.
With such an initialization scheme, these methods are at risk of converging to a poor local minima.
\emph{\ouralgorithm} overcomes this problem by introducing a multilevel paradigm for graph representation learning.

\noindent
\textbf{Graph Drawing}.
Multilevel layout algorithms are popular methods in the graph drawing community, where a hierarchy of approximations is used to solve the original layout problem
\cite{fruchterman1991graph,hu2005efficient,walshaw2003multilevel}. 
Using an approximation of the original graph has two advantages - not only is the approximation usually simpler to solve, it can also be extended as a good initialization for solving the original problem. 
In addition to force-directed graph drawing, the multilevel framework \cite{walshaw2004multilevel} has been proved successful in various graph theory problems, including the traveling salesman problem \cite{walshaw2001multilevel}, and graph partitioning \cite{karypis1998parallel}.

\emph{\ouralgorithm} extends the idea of the multilevel layout to neural representation learning methods.
We illustrate the utility of this paradigm by combining \emph{\ouralgorithm} with three state-of-the-art representation learning methods.

\section{Conclusion}
\label{Conclusion}

Recent literature on graph representation learning aims at optimizing a non-convex function.
With no prior knowledge of the graph, these methods could easily get stuck at a bad local minima as the result of poor initialization.
Moreover, these methods mostly aim to preserve local proximities in a graph but neglect its global structure.
In this paper, we propose a multilevel graph representation learning paradigm to address these issues.
By recursively coalescing the input graph into smaller but structurally similar graphs, \emph{\ouralgorithm} captures the global structure of the input graph.
By learning graph representation on these smaller graphs, a good initialization scheme for the input graph is derived.
This multilevel paradigm is further combined with the state-of-the-art graph embedding methods, namely \emph{DeepWalk}, \emph{LINE}, and \emph{Node2vec}. Experimental results on various real-world graphs show that introducing \emph{\ouralgorithm} yields graph embeddings of higher quality for all these three methods.

In the future, we would like to combine \emph{\ouralgorithm} with other graph representation learning methods.
Specifically, as Skip-gram is a shallow method for representation learning, it would be interesting to see if \emph{\ouralgorithm} also works well with deep representation learning methods.
On the other hand, our method could also be applied to language networks, possibly yielding better word embeddings.

\section{Acknowledgements}
\label{Ackonwledgements}
This work is partially supported by NSF grants IIS-1546113 and DBI-1355990.

\bibliography{chen-perozzi}

\begin{thebibliography}{}

\bibitem[\protect\citeauthoryear{Abu-El-Haija, Perozzi, and
  Al-Rfou}{2017}]{abu2017learning}
Abu-El-Haija, S.; Perozzi, B.; and Al-Rfou, R.
\newblock 2017.
\newblock Learning edge representations via low-rank asymmetric projections.
\newblock {\em arXiv preprint arXiv:1705.05615}.

\bibitem[\protect\citeauthoryear{Belkin and Niyogi}{2001}]{belkin2001laplacian}
Belkin, M., and Niyogi, P.
\newblock 2001.
\newblock Laplacian eigenmaps and spectral techniques for embedding and
  clustering.
\newblock In {\em NIPS}, volume~14,  585--591.

\bibitem[\protect\citeauthoryear{Cao, Lu, and Xu}{2015}]{cao2015grarep}
Cao, S.; Lu, W.; and Xu, Q.
\newblock 2015.
\newblock Grarep: Learning graph representations with global structural
  information.
\newblock In {\em Proceedings of the 24th ACM International on Conference on
  Information and Knowledge Management},  891--900.
\newblock ACM.

\bibitem[\protect\citeauthoryear{Cox and Cox}{2000}]{cox2000multidimensional}
Cox, T.~F., and Cox, M.~A.
\newblock 2000.
\newblock {\em Multidimensional scaling}.
\newblock CRC press.

\bibitem[\protect\citeauthoryear{Fan \bgroup et al\mbox.\egroup
  }{2008}]{fan2008liblinear}
Fan, R.-E.; Chang, K.-W.; Hsieh, C.-J.; Wang, X.-R.; and Lin, C.-J.
\newblock 2008.
\newblock Liblinear: A library for large linear classification.
\newblock {\em The Journal of Machine Learning Research} 9:1871--1874.

\bibitem[\protect\citeauthoryear{Fruchterman and
  Reingold}{1991}]{fruchterman1991graph}
Fruchterman, T.~M., and Reingold, E.~M.
\newblock 1991.
\newblock Graph drawing by force-directed placement.
\newblock {\em Software: Practice and experience} 21(11):1129--1164.

\bibitem[\protect\citeauthoryear{Goldberg and
  Levy}{2014}]{goldberg2014word2vec}
Goldberg, Y., and Levy, O.
\newblock 2014.
\newblock word2vec explained: deriving mikolov et al.'s negative-sampling
  word-embedding method.
\newblock {\em arXiv preprint arXiv:1402.3722}.

\bibitem[\protect\citeauthoryear{Grover and Leskovec}{2016}]{node2vec-kdd2016}
Grover, A., and Leskovec, J.
\newblock 2016.
\newblock node2vec: Scalable feature learning for networks.
\newblock In {\em Proceedings of the 22nd ACM SIGKDD International Conference
  on Knowledge Discovery and Data Mining}.

\bibitem[\protect\citeauthoryear{Hu}{2005}]{hu2005efficient}
Hu, Y.
\newblock 2005.
\newblock Efficient, high-quality force-directed graph drawing.
\newblock {\em Mathematica Journal} 10(1):37--71.

\bibitem[\protect\citeauthoryear{Karypis and Kumar}{1998}]{karypis1998parallel}
Karypis, G., and Kumar, V.
\newblock 1998.
\newblock A parallel algorithm for multilevel graph partitioning and sparse
  matrix ordering.
\newblock {\em Journal of Parallel and Distributed Computing} 48(1):71--95.

\bibitem[\protect\citeauthoryear{Mikolov \bgroup et al\mbox.\egroup
  }{2013}]{mikolov2013distributed}
Mikolov, T.; Sutskever, I.; Chen, K.; Corrado, G.~S.; and Dean, J.
\newblock 2013.
\newblock Distributed representations of words and phrases and their
  compositionality.
\newblock In {\em Advances in neural information processing systems},
  3111--3119.

\bibitem[\protect\citeauthoryear{Ou \bgroup et al\mbox.\egroup
  }{2016}]{ouasymmetric}
Ou, M.; Cui, P.; Pei, J.; and Zhu, W.
\newblock 2016.
\newblock Asymmetric transitivity preserving graph embedding.
\newblock In {\em Proceedings of the 22nd ACM SIGKDD International Conference
  on Knowledge Discovery and Data Mining}.

\bibitem[\protect\citeauthoryear{Perozzi, Al-Rfou, and
  Skiena}{2014}]{perozzi2014deepwalk}
Perozzi, B.; Al-Rfou, R.; and Skiena, S.
\newblock 2014.
\newblock Deepwalk: Online learning of social representations.
\newblock In {\em Proceedings of the 20th ACM SIGKDD international conference
  on Knowledge discovery and data mining},  701--710.
\newblock ACM.

\bibitem[\protect\citeauthoryear{Perozzi \bgroup et al\mbox.\egroup
  }{2017}]{walklets}
Perozzi, B.; Kulkarni, V.; Chen, H.; and Skiena, S.
\newblock 2017.
\newblock Don't walk, skip!: Online learning of multi-scale network embeddings.
\newblock In {\em Proceedings of the 2017 IEEE/ACM International Conference on
  Advances in Social Networks Analysis and Mining 2017}, ASONAM '17,  258--265.
\newblock New York, NY, USA: ACM.

\bibitem[\protect\citeauthoryear{Roweis and Saul}{2000}]{roweis2000nonlinear}
Roweis, S.~T., and Saul, L.~K.
\newblock 2000.
\newblock Nonlinear dimensionality reduction by locally linear embedding.
\newblock {\em Science} 290(5500):2323--2326.

\bibitem[\protect\citeauthoryear{Sen \bgroup et al\mbox.\egroup
  }{2008}]{sen:aimag08}
Sen, P.; Namata, G.~M.; Bilgic, M.; Getoor, L.; Gallagher, B.; and Eliassi-Rad,
  T.
\newblock 2008.
\newblock Collective classification in network data.
\newblock {\em AI Magazine} 29(3):93--106.

\bibitem[\protect\citeauthoryear{Tang and Liu}{2009}]{tang2009relational}
Tang, L., and Liu, H.
\newblock 2009.
\newblock Relational learning via latent social dimensions.
\newblock In {\em Proceedings of the 15th ACM SIGKDD international conference
  on Knowledge discovery and data mining},  817--826.
\newblock ACM.

\bibitem[\protect\citeauthoryear{Tang \bgroup et al\mbox.\egroup
  }{2015}]{tang2015line}
Tang, J.; Qu, M.; Wang, M.; Zhang, M.; Yan, J.; and Mei, Q.
\newblock 2015.
\newblock Line: Large-scale information network embedding.
\newblock In {\em Proceedings of the 24th International Conference on World
  Wide Web},  1067--1077.
\newblock International World Wide Web Conferences Steering Committee.

\bibitem[\protect\citeauthoryear{Tenenbaum, De~Silva, and
  Langford}{2000}]{tenenbaum2000global}
Tenenbaum, J.~B.; De~Silva, V.; and Langford, J.~C.
\newblock 2000.
\newblock A global geometric framework for nonlinear dimensionality reduction.
\newblock {\em Science} 290(5500):2319--2323.

\bibitem[\protect\citeauthoryear{Walshaw}{2001}]{walshaw2001multilevel}
Walshaw, C.
\newblock 2001.
\newblock {\em A multilevel Lin-Kernighan-Helsgaun algorithm for the travelling
  salesman problem}.
\newblock Citeseer.

\bibitem[\protect\citeauthoryear{Walshaw}{2003}]{walshaw2003multilevel}
Walshaw, C.
\newblock 2003.
\newblock A multilevel algorithm for force-directed graph-drawing.
\newblock {\em Journal of Graph Algorithms Applications} 7(3):253--285.

\bibitem[\protect\citeauthoryear{Walshaw}{2004}]{walshaw2004multilevel}
Walshaw, C.
\newblock 2004.
\newblock Multilevel refinement for combinatorial optimisation problems.
\newblock {\em Annals of Operations Research} 131(1-4):325--372.

\bibitem[\protect\citeauthoryear{Wang, Cui, and Zhu}{2016}]{wangstructural}
Wang, D.; Cui, P.; and Zhu, W.
\newblock 2016.
\newblock Structural deep network embedding.
\newblock In {\em Proceedings of the 22nd ACM SIGKDD International Conference
  on Knowledge Discovery and Data Mining}.

\end{thebibliography}
\bibliographystyle{aaai}

\end{document}